\documentclass[a4paper,12pt]{article}

\usepackage[english]{babel}
\usepackage{color}
\usepackage{latexsym}
\usepackage{
euscript, mathrsfs}
\usepackage{braket}
\usepackage{mathtools}
\usepackage{amsmath}
\usepackage{bbm} 
\usepackage{verbatim}

\usepackage{jheppub}                   
\makeatletter
\gdef\@fpheader{\ }                    
\makeatother

\DeclarePairedDelimiter{\abs}{\lvert}{\rvert}

\newcommand{\CO}{\mathcal{O}}
\newcommand{\CN}{\mathcal{N}}

\newcommand{\CV}{\mathcal{V}}
\newcommand{\CS}{\mathcal{S}}

\newcommand{\CR}{\mathcal{R}}
\newcommand{\CW}{\mathcal{W}}

\def\be{\begin{equation}}
\def\ee{\end{equation}}
\def\bea{\begin{eqnarray}}
\def\eea{\end{eqnarray}}
\def\rme{{\rm e}}
\newcommand{\nn}{\nonumber}

\newcommand{\diff}{\mathrm{d}}

\newcommand{\SO}{{\rm SO}}
\newcommand{\SU}{{\rm SU}}

\newcommand{\U}{{\rm U}}

\def\rnot{\bar{r}}
\def\scal{\xi}

\title{The BPS limit of rotating AdS black hole thermodynamics}

\author[a]{Davide Cassani,}
\emailAdd{davide.cassani@pd.infn.it}
\author[\,a,b]{Lorenzo Papini}
\emailAdd{lorenzo.papini@pd.infn.it}

\affiliation[a]{INFN, Sezione di Padova, Via Marzolo 8, 35131 Padova, Italy}

\affiliation[b]{Dipartimento di Fisica e Astronomia ``Galileo Galilei'', Via Marzolo 8, 35131 Padova, Italy}

\abstract{
We consider rotating, electrically charged, supersymmetric AdS black holes in four, five, six and seven dimensions, and provide a derivation of the respective extremization principles stating that the Bekenstein-Hawking entropy is the Legendre transform of a homogeneous function of chemical potentials, subject to a complex constraint.
Extending a recently proposed BPS limit, we start from finite temperature and reach extremality following a supersymmetric trajectory in the space of complexified solutions. We show that the  entropy function is the supergravity on-shell action in this limit. Chemical potentials satisfying the extremization equations also emerge from the complexified solution.
}

\begin{document}
\maketitle

\section{Introduction}\label{sec:introduction}

In a fundamental theory of quantum gravity, the thermodynamic properties of black holes should emerge from a microscopic statistical description. A major achievement of string theory has been to provide precisely the microstates accounting for the entropy of certain classes of supersymmetric black holes \cite{Strominger:1996sh}. 
While most of the original work focused on asymptotically flat black holes, in the last few years there has been a lot of progress on black holes in Anti de Sitter (AdS) space and their holographic dual conformal field theories (CFT's), starting with \cite{Benini:2015eyy,Benini:2016rke} (see \cite{Zaffaroni:2019dhb} for a review and a comprehensive list of references). In this paper we will be interested in supersymmetric, asymptotically AdS black holes with non-trivial rotation, where very recent progress has been made in \cite{Cabo-Bizet:2018ehj,Choi:2018hmj,Benini:2018ywd,Honda:2019cio,ArabiArdehali:2019tdm,Choi:2019miv,Kim:2019yrz,Cabo-Bizet:2019osg,Amariti:2019mgp}.

The main reason for considering asymptotically AdS spacetimes is that quantum gravity in such spaces can be understood in terms of a dual CFT via the AdS/CFT correspondence, and this makes it in principle possible to provide a complete characterization of the black hole microstates.\footnote{For extremal black holes (irrespective of asymptotics) one can also isolate an AdS$_2$ region by zooming in on the near horizon. Then one can use the AdS$_2$/CFT$_1$ correspondence to formulate the entropy problem at the full quantum level, following Sen's approach~\cite{Sen:2008yk}. However the relevant CFT$_1$ is not known in general. We will comment on the relation of our asymptotically AdS analysis with Sen's approach later on in this section.}
 Conversely, studying black hole solutions to string theory teaches us about the statistical behavior of interesting ensembles of states in holographic CFT's.
For both sides of the correspondence to be under good control, one typically demands some supersymmetry to be preserved.

A general feature that has been identified in the last few years is that the Bekenstein-Hawking entropy of BPS black holes in AdS arises from an {\it extremization principle} \cite{Benini:2015eyy,Benini:2016rke,Hosseini:2017mds,Hosseini:2018dob,Choi:2018fdc}. The entropy is indeed the Legendre transform of a rather simple, homogeneous {\it entropy function} of rotational and electric chemical potentials $\omega^i,\varphi^I$, subject to a linear, complex constraint that schematically reads
\be\label{constraint_intro}
\sum_i \omega^i - \sum_I \varphi^I =  2\,\pi\, i\ .
\ee
Since field theory partition functions are more directly computed as a function of  chemical potentials (that is background fields) rather than charges (that is expectation values), it is convenient to recast the problem of the black hole entropy into the one of computing the corresponding entropy function. Of course, this is just a change of statistical ensemble. It is thus interesting to ask what is the physical interpretation of the entropy function, both on the gravity and the field theory side of the holographic correspondence.
In the case of rotating black holes, the question is subtle because the saddle point values of both the rotational and electric chemical potentials turn out to be complex, as enforced by the constraint \eqref{constraint_intro}. Although this may not be surprising if one recalls that rotating spacetimes are related to complex saddles of the gravitational path integral \cite{Gibbons:1976ue,Brown:1990fk}, it is not obvious how to read the chemical potentials in \eqref{constraint_intro} from the black hole solution.
These issues were solved in \cite{Cabo-Bizet:2018ehj}, where it was found that the entropy function for a class  of rotating BPS black holes in AdS$_5$ is the supergravity on-shell action after taking a specific BPS limit, that goes along a supersymmetric trajectory in the space of complexified solutions.
 Further, the constraint \eqref{constraint_intro} was interpreted as a regularity condition for the Killing spinor of the supersymmetric solution, which is anti-periodic when going around the compactified Euclidean time in the smooth cigar-like geometry.\footnote{The entropy function of non-rotating BPS black holes has been related to the supergravity on-shell action in \cite{Halmagyi:2017hmw,Azzurli:2017kxo,Cabo-Bizet:2017xdr} for AdS$_4$ black holes and in \cite{Suh:2018qyv} for AdS$_6$ black holes.}

In this paper, we extend the analysis of \cite{Cabo-Bizet:2018ehj} to other classes of rotating, asymptotically AdS black holes in different dimensions. While the five-dimensional black hole discussed in \cite{Cabo-Bizet:2018ehj} 
carries two angular momenta and one electric charge (dual to the R-charge of generic four-dimensional $\mathcal{N}=1$ superconformal field theories), here we discuss the case with one angular momentum and multiple electric charges. We also analyze four-, six- and seven-dimensional black holes.
 The solutions we consider were presented in \cite{Cvetic:2005zi,Chow:2008ip}. In addition to the Bekenstein-Hawking entropy $S$, these carry macroscopic energy $E$, angular momenta $J_i$ and electric charges $Q_I$ (with some of the possible angular momenta or electric charges being set equal in the explicit solutions of \cite{Cvetic:2005zi,Chow:2008ip}). Conjugate to the charges $E,J_i,Q_I$ one has the chemical potentials $\beta,\Omega^i,\Phi^I$, where $\beta$ is the inverse Hawking temperature, $\Omega^i$ are the angular velocities and $\Phi^I$ are the electrostatic potentials of the black hole. The corresponding expressions can be read off from the solution by standard methods \cite{Cvetic:2005zi}. In five and four dimensions, we explicitly evaluate the Euclidean action $I$ of the finite-temperature, non-supersymmetric solution using holographic renormalization, and verify that it satisfies the {\it quantum statistical relation}
\begin{equation}\label{QSR_nonsusy}
I = \beta E -  S - \beta\,\Omega^i  J_i - \beta\,\Phi^I  Q_I  \, .
\end{equation}
This well-known relation, first proposed for quantum gravity in \cite{Gibbons:1976ue} (see e.g.~\cite{Papadimitriou:2005ii} for a discussion in relation with holography) is expected to hold in full generality, and we assume that it is also satisfied in six and seven dimensions. The quantum statistical relation makes it manifest that the on-shell action has an interpretation as a thermodynamic potential. From a microscopic point of view, the latter corresponds to minus the logarithm of the grand-canonical partition function, while the entropy is the logarithm of the microcanonical partition function.

Starting from the non-supersymmetric and non-extremal solution, we want to reach the BPS locus in parameter space, namely the solution that is both supersymmetric and extremal.
Motivated by the fact that in the dual field theory one is mostly interested in studying a supersymmetric ensemble of states, we adopt the strategy of~\cite{Cabo-Bizet:2018ehj} and first impose supersymmetry, namely that the supergravity Killing spinor equations are solved. This amounts to precisely one condition on the parameters of the original solution, and it is important to remark that it does not automatically imply vanishing of the temperature.\footnote{For this reason, throughout the paper we will carefully distinguish between  supersymmetry and extremality. A quantity evaluated after imposing both supersymmetry and extremality will be called ``BPS'' and denoted by the symbol $\star$ in the formulae. For instance, $S^\star$ is the BPS entropy.
} In Lorentzian signature, this supersymmetric solution has causal pathologies unless one also sends the temperature to zero \cite{Cvetic:2005zi}.  
 However, here we are interested in semi-classical saddle points of the Euclidean path integral, and thus allow for more general solutions by complexifying (one of) the remaining parameters.  
Further, building on ideas of~\cite{Silva:2006xv}, we introduce the variables
\be\label{smallpots_from_bogpots}
\omega^i = \beta\left(\Omega^i-\Omega^{i\,\star}\right)\,,\qquad \varphi^I = \beta\,(\Phi^I - \Phi^{I\star})\,,
\ee
 where $\Omega^{i\,\star}$ and $\Phi^{I\,\star}$ are the (frozen) values taken by $\Omega^i$ and $\Phi^I$ in the BPS solution. 
 The variables \eqref{smallpots_from_bogpots} are the chemical potentials conjugate to the angular momentum and electric charges when one identifies the generator of ``time'' translations with the conserved quantity  $ \{\mathcal{Q},\overline{\mathcal{Q}}\}$, where $\mathcal{Q}$ is the supercharge.\footnote{In odd dimensions, there are some subtleties with the regularization of a priori divergent quantities that we will address in Section~\ref{5d_section}.} 
The asymptotic analysis of the supergravity solution defines the dual superconformal field theory (SCFT) partition function, $Z$. From the previous considerations, one infers the following Hamiltonian representation:
\be\label{SCFT_Z}
Z = {\rm Tr}\big[\rme^{-\beta\{\mathcal{Q},\overline{\mathcal{Q}}\} +\omega^iJ_i + \varphi^I Q_I} \big]\,,
\ee
where there is no $(-1)^F$ due to anti-periodicity of the supercharge. It was shown in~\cite{Cabo-Bizet:2018ehj} that upon using \eqref{constraint_intro}, $Z$ is proportional to the superconformal index \cite{Romelsberger:2005eg,Kinney:2005ej}. Note that we have an identification between the SCFT chemical potentials appearing in \eqref{SCFT_Z} and the black hole variables \eqref{smallpots_from_bogpots}.
 
For each of the cases that we analyze, we verify that after imposing supersymmetry the variables \eqref{smallpots_from_bogpots} satisfy a linear relation of the type \eqref{constraint_intro}, and are otherwise free. Moreover, the supersymmetric on-shell action $I$ takes the form of a simple function of these variables, that precisely matches the entropy functions proposed in \cite{Hosseini:2017mds,Hosseini:2018dob,Choi:2018fdc}. The supersymmetric form of the quantum statistical relation \eqref{QSR_nonsusy} is
\begin{equation}
I = - S -  \omega^i J_i -  \varphi^I \, Q_I   	\, ,
\end{equation}
and the first law of thermodynamics in the supersymmetric ensemble reads
\be
\diff S + \omega^i\, \diff J_i + \varphi^I \, \diff Q_I = 0\,.
\ee
The Legendre transform of $I$ (subject to the constraint \eqref{constraint_intro}) is in general a complex quantity, so it cannot be immediately identified with the entropy of the Lorentzian solution. However, demanding reality of the Legendre transform, which amounts to a specific condition on the charges,  one finds precisely the Bekenstein-Hawking entropy of the supersymmetric {\it and extremal} black hole~\cite{Cabo-Bizet:2018ehj,Choi:2018hmj}. The saddle point values of the chemical potentials remain complex and match the ones that we obtain from the solution by taking the zero-temperature limit of \eqref{smallpots_from_bogpots}. 
%
We thus conclude that the BPS limit of black hole thermodynamics described above gives a derivation of the proposed entropy functions and the related extremization principles, based on black hole thermodynamics and semiclassical Euclidean quantum gravity. This extends the results of \cite{Cabo-Bizet:2018ehj} and shows that the procedure presented there has general validity.
 It is important to emphasize that the on-shell action entering in the quantum statistical relation is first defined in a regular Euclidean solution where the Wick-rotated time has been compactified and the metric is positive-definite. After the on-shell action has been computed in this way, one can extend its value to a complexified solution by analytic continuation~\cite{Gibbons:1976ue}. As it should be clear from the discussion above, we find that this complexification is crucial for the on-shell action to eventually match the proposed entropy functions.

We observe that for the same solutions of \cite{Cvetic:2005zi,Chow:2008ip} studied in this paper, the BPS limit of the quantum statistical relation was also considered in \cite{Choi:2018hmj,Choi:2018fdc}. However our limit is different precisely because it reaches the physical BPS black hole through a complexified family of solutions, specified by supersymmetry. The limit taken in \cite{Choi:2018hmj,Choi:2018fdc}  appears similar to the one originally discussed in \cite{Silva:2006xv}, in that it yields real chemical potentials that satisfy just the real part of \eqref{constraint_intro}. Correspondingly, the on-shell action in this limit does not match the entropy functions proposed in \cite{Hosseini:2017mds,Hosseini:2018dob,Choi:2018fdc}.

A related comment concerns the connection of our approach with Sen's entropy function formalism, which is based on a near-horizon analysis of extremal black holes (regardless of supersymmetry) \cite{Sen:2007qy,Sen:2008yk}. Sen's formalism leads to real chemical potentials and a real entropy function. Correspondingly, it was shown in~\cite{Dias:2007dj} that this is matched by an extremal limit of black hole thermodynamics where all quantities remain real. 
  As discussed in~\cite[Sect.$\:$3.3]{Cabo-Bizet:2018ehj}, there is in fact a continuous family of extremal limits of black hole thermodynamics, all leading to a meaningful entropy function and an associated constraint between chemical potentials. However only the manifestly supersymmetric limit discussed above leads to the entropy functions proposed in \cite{Hosseini:2017mds,Hosseini:2018dob,Choi:2018fdc}. It would  be interesting to investigate further the relation between this supersymmetric limit and Sen's near-horizon approach.

The rest of the paper is organized as follows. We start in Section~\ref{5d_section} by studying the five-dimensional solutions and making contact with the results of \cite{Cabo-Bizet:2018ehj}. We continue in Section \ref{4d_section} by analyzing the four-dimensional solutions, while in Section~\ref{6d_section} we turn to six dimensions and in Section~\ref{7d_section} we move to seven dimensions. In each case we start with a brief review of the finite-temperature, non-supersymmetric solution, and then discuss our BPS limit.     
 We conclude in Section~\ref{sec:conclusions}.
Appendix~\ref{app:OnShActionsAdS5} and Appendix~\ref{app:OnShActionsAdS4} contain the details on the computation of the on-shell action in five and four dimensions, respectively, while Appendix \ref{app:Leg_transf} describes the Legendre transformation reproducing the Bekenstein-Hawking entropy of the general BPS black hole solutions to five-dimensional Fayet-Iliopoulos gauged supergravity constructed in \cite{Gutowski:2004yv,Kunduri:2006ek}.

\section{Rotating AdS$_5$ black holes with multiple electric charges}\label{5d_section}

In this section we study the BPS limit of AdS$_5$ black hole thermodynamics. 
In \cite{Cabo-Bizet:2018ehj} a new BPS limiting procedure was defined and applied to the solution of \cite{Chong:2005hr}, which has two independent angular momenta and ---being constructed within  five-dimensional minimal gauged supergravity--- just one electric charge. 
Here we discuss a different setup, including multiple electric charges.
 We will focus on solutions to  $\U(1)^3$ Fayet-Iliopoulos gauged supergravity, that is $\mathcal{N}=2$ supergravity coupled to two vector multiplets and with a $\U(1)$ gauging of the R-symmetry. This theory uplifts to type IIB supergravity on $S^5$, hence the dual SCFT is $\mathcal{N}=4$ SYM or an orbifold thereof.
The most general set of black hole conserved charges in the theory is given by the energy, three electric charges and two angular momenta.
 When imposing supersymmetry and extremality these satisfy two relations, hence the BPS solution carries four independent conserved charges \cite{Kunduri:2006ek}. Here we will discuss a solution where the two angular momenta are set equal while the three electric charges are independent, originally found in \cite{Cvetic:2004ny} and further discussed in \cite{Cvetic:2005zi}. This solution contains the three-parameter BPS black hole of \cite{Gutowski:2004yv}.\footnote{The solution where all the conserved charges  are independent was given in \cite{Wu:2011gq}. Solutions with restricted set of independent charges were also found in \cite{Chong:2005da,Chong:2006zx,Mei:2007bn}.} 

We first check the finite-temperature quantum statistical relation by explicitly computing the Euclidean on-shell action using holographic renormalization.
Then we impose supersymmetry and obtain complex chemical potentials satisfying the constraint \eqref{constraint_intro} (with one rotational chemical potential $\omega$). We further find that the supersymmetric on-shell action reproduces the entropy function. Then we show that these properties are preserved in the BPS limit.

\subsection{The non-supersymmetric finite-temperature solution}\label{sec:nonsusy_5d_sol}

We start by briefly reviewing the non-supersymmetric, finite temperature solution of \cite{Cvetic:2004ny}, mostly following the presentation of \cite{Cvetic:2005zi}.
The five-dimensional action is\footnote{In this paper we set to 1 the Newton constant $G$.}
\begin{align}
\CS = \frac{1}{16\pi}\int &\Big[\Big(R  + 4  g^2 \sum^3_{I=1} \left(X^I \right)^{-1} - \frac{1}{2}  \partial \vec{\phi}^{\,2} \Big)\star1 - \frac{1}{2} \sum_{I=1}^3  \left(X^I\right)^{-2} F^I \wedge \star F^I  \nn\\[1mm]
& \ - \frac{1}{6}\,  |\epsilon_{IJK}| \, A^I \wedge F^J \wedge F^K \Big] \,, \label{CGLP_Lagrangian}
\end{align} 
where $A^I$, $I=1,2,3,$ are Abelian gauge fields, with field strength $F^I =\diff A^I$, while $\vec{\phi} = \left(\phi_1, \phi_2 \right)$ are real scalar fields and
\begin{equation}
X^1 = \rme^{- \frac{1}{\sqrt{6}} \phi_1 - \frac{1}{\sqrt{2}} \phi_2} , \qquad X^2 = \rme^{- \frac{1}{\sqrt{6}} \phi_1 + \frac{1}{\sqrt{2}} \phi_2} , \qquad X^3 = \rme^{\frac{2}{\sqrt{6}} \, \phi_1}\,.
\end{equation}
This theory is a consistent truncation of type IIB supergravity on $S^5$, where the $A^I$ arise as Kaluza-Klein vector fields gauging the $\U(1)^3\subset\SO(6)$ isometries of $S^5$ \cite{Cvetic:1999xp}. It is also a consistent truncation of five-dimensional maximal $\SO(6)$ supergravity.
We describe how it fits in the framework of Fayet-Iliopoulos gauged $\mathcal{N}=2$ supergravity in  Appendix~\ref{app:OnShActionsAdS5}.

The solution is expressed in terms of coordinates $(t,r,\theta,\phi,\psi)$ and uses the following left-invariant 1-forms on a three-sphere $S^3$ parameterized by $(\theta,\phi,\psi)$:
\begin{align}
\label{Sigma_1_forms}
\sigma_1+ i\, \sigma_2 &=  \rme^{-i\psi}(\diff \theta +i \sin\theta\,\diff\phi) \, ,\notag \\
 \sigma_3 &= \diff  \psi + \cos \theta \, \diff  \phi \, .
\end{align}
These will make $\SU(2) \times \U(1)$ symmetry manifest.
The metric, scalar fields and gauge fields read:
\begin{align}
\label{Metric_CLP}
\diff s^2_5 &= \left(H_1  H_2 H_3 \right)^{1/3} \!\left[ - \frac{ r^2 \,Y }{f_1}  \diff t^2 + \frac{r^4 }{Y}  \diff r^2 + \frac{r^2}{4} \big(\sigma^2_1 + \sigma^2_2 \big) + \frac{f_1}{4 r^4 H_1  H_2  H_3 } \Big(\sigma_3 - \frac{2f_2}{f_1}  \diff t \Big)^2  \right]\! , \\
X^I &= \frac{ \left(H_1 H_2  H_3 \right)^{1/3}} {H_I} \, ,\label{Scalar_form} \\[1mm]
A^I &= A^I_t \, \diff t + A^I_\psi \, \sigma_3 \, ,\label{Gauge_Fields}
\end{align}
where
\be
 A^I_t = \frac{2 \, m}{r^2 \, H_I} \, s_I \, c_I + \alpha^I \, ,\label{time_component_A}\ \qquad
 A^I_\psi = \frac{m \, a}{r^2 \, H_I} \, \left(c_I \, s_J \, s_K - s_I \, c_J \, c_K \right) \, ,
\ee
and the indices $I,J,K$ in $A^I_\psi$ are never equal. 
Here we introduced the following functions of the radial coordinate $r$:
\begin{align}\label{functions_5d}
& H_I = 1 + \frac{2 \, m \, s_I^2}{r^2} \, ,\notag \\[1mm]
& f_1 = r^6H_1  H_2 H_3  + 2\, m \, a^2 r^2 + 4 \, m^2  a^2 \left[ 2 \left(c_1  c_2  c_3 - s_1  s_2  s_3 \right)  s_1  s_2  s_3 - s_1^2  s_2^2 - s_2^2 s_3^2 -  s_3^2s_1^2 \right] , \notag  \\[1mm]
& f_2 = 
 2 \, m \, a \left(c_1  c_2  c_3 - s_1 s_2 s_3 \right) \, r^2 + 4 \, m^2  a \, s_1  s_2  s_3 \, , \notag \\[1mm]
& f_3 = 
   2 \, m \, a^2 (1+ g^2  r^2)  + 4 \, g^2  m^2  a^2 \left[ 2 (c_1  c_2  c_3 - s_1  s_2  s_3 )  s_1 s_2 s_3 -  s_1^2  s_2^2 - s_2^2 s_3^2 - s_3^2 s_1^2  \right]   , \notag \\[1mm]
& Y = f_3 + g^2 r^6 H_1  H_2 H_3 + r^4 - 2 \, m \, r^2  \, ,
\end{align}
and $s_I$, $c_I$ are shorthand notations for:
\begin{equation}
s_I = \sinh{\delta_I} \, , \qquad c_I = \cosh{\delta_I} \, , \qquad I = 1,2,3 \, .
\end{equation}

The solution depends on the five parameters
$m , \, \delta_1 , \, \delta_2  , \, \delta_3  , \, a$. In the temporal component of the gauge fields we also introduced a constant gauge choice $\alpha^I$, that will be fixed soon. The five parameters should satisfy suitable inequalities, so that the spatial components of the metric are positive for $r>r_+$, where $r_+$ denotes the position of the outer horizon, given by the largest positive root of $Y(r)$. This is a Killing horizon since the Killing vector
\begin{equation}
\label{Killing_Vector}
V = \frac{\partial}{\partial t} + 2 \,\frac{ f_2 (r_+)}{f_1 (r_+)} \, \frac{\partial}{\partial \psi} \, 
\end{equation}
is null at $r=r_+$. To the outer event horizon we can associate the quantities:
\begin{align}
\label{Entropy_Temperature_Potentials}
& S = \frac{\pi^2 }{2} \, \sqrt{f_1 (r_+)} \, , \qquad\,  \beta = 4 \, \pi \, r_+ \sqrt{f_1 (r_+)} \left( \frac{\diff Y}{\diff r} (r_+)\right)^{-1}\, ,  \notag \\
&\Omega = 2 \,\frac{ f_2 (r_+ ) }{f_1 (r_+ )} \,,\quad \qquad  \Phi^I = \frac{2 \, m}{r_+^2 \, H_I(r_+)} \left( s_I \, c_I + \tfrac{1}{2} \, a \,  \Omega \left(c_I \, s_J \, s_K - s_I \, c_J \, c_K  \right) \right)\,, 
\end{align}
where $S$ is the Bekenstein-Hawking entropy computed as $\frac{1}{4}$ the area of the horizon, $\beta = T^{-1}=\frac{2\pi}{\kappa}$ is the inverse Hawking temperature obtained from the surface gravity $\kappa$, $\Omega$ is the angular velocity relative to a non-rotating frame at infinity as read off from the Killing vector $V$, and $\Phi^I$ are the electrostatic potentials,\footnote{We corrected a minus sign typo in the expression for $\Phi^I$ given in eq.~(3.10) of~\cite{Cvetic:2005zi}, see also~\cite{Choi:2018hmj}.} defined as 
\be
\Phi^I=\iota_V A^I|_{r_+} - \iota_V A^I|_\infty\,.
\ee

Corresponding to the five parameters there are five independent conserved charges. These are the energy $E$ for translations along $\frac{\partial}{\partial t}$, the angular momentum $J$ for rotations along $-\frac{\partial}{\partial\psi}$, and three electric charges $Q_I$. Their values are:
\begin{align}
\label{nonextremal_charges}
E &= E_0 + \frac{1}{4} \, m \, \pi \left(3 + a^2 g^2 + 2 \, s_1^2 + 2 \, s_2^2 + 2 \, s_3^2 \right) \,,\nn \\[1mm]
J &= \frac{1}{2} \, m \, a \, \pi \, \left(c_1 \, c_2 \, c_3 - s_1 \, s_2 \, s_3 \right) \, ,\nn \\[1mm]
Q_I &= \frac{1}{2} \, m \, \pi \, s_I \, c_I \, .
\end{align}
In \cite{Cvetic:2005zi}, the electric charges and the angular momentum above were computed using the boundary integrals\footnote{In the formula for the electric charge, we omitted the contribution from the Chern-Simons term in the action, since this vanishes in the solution of interest (because $F^I\to0$ as $r\to\infty$). This implies that a priori different definitions of the electric charge such as the Maxwell charge, the Page charge, and the charge that arises from integrating the holographic current, actually coincide in the present background.
}
\begin{align}
Q_I &= - \frac{1}{16  \pi} \int_{S^3_{\rm bdry}} {   \left(X^I \right)^{-2}  \star F^I  } \, ,\nn\\[1mm]
J &= \frac{1}{16  \pi} \int_{S^3_{\rm bdry}} \star\, \diff \left(g_{\psi \mu} \diff x^\mu\right) \ ,\label{QMaxwell_JKomar}
\end{align}
while the energy $E$ was obtained by integrating the first law of thermodynamics,
\begin{equation}\label{first_law_5d}
\diff E = T \, \diff S + \Omega \, \diff J +  \Phi^I \, \diff Q_I 
 \, .
\end{equation}
 The integration constant $E_0$ was fixed to zero in \cite{Cvetic:2005zi} by requiring that $E$ vanishes in the limiting case $m=0$ where the solution becomes empty AdS$_5$, which is regarded as the vacuum solution (see \cite{Gibbons:2004ai} for more details on this approach to computing the energy).

One can also compute the same charges within the framework of holographic renormalization. We do so  in Appendix~\ref{app:OnShActionsAdS5},  adopting  a minimal subtraction scheme. As expected from the analysis of \cite{Papadimitriou:2005ii}, we find agreement with the expressions above for the angular momentum $J$ and the electric charges $Q_I$. The energy $E$  also agrees, except that the AdS mass $E_0$ now takes the non-vanishing value
\be\label{eq:E0}
E_0 = \frac{3 \, \pi}{32 \, g^2}\,.
\ee

The on-shell action of this solution does not appear to have been computed in the literature before. We have done so, again using holographic renormalization. 
The action must be evaluated on a regular Euclidean section of the solution. The Euclideanization is obtained by the Wick rotation $t=-i\tau$, together with the continuation of the parameter $a$ to purely imaginary values. After the action is computed one can take $a$ back to the original real domain, or choose to analytically continue the solution to more general complex values of the parameters \cite{Gibbons:1976ue}.
As usual, regularity of the Euclidean section leads to identify the length of the circle parameterized by the Euclidean time $\tau $ with the inverse Hawking temperature, that is $\int \diff\tau=\beta$.
 A further regularity condition is that the contraction of the Killing vector \eqref{Killing_Vector} with the gauge fields vanishes at the horizon,
\begin{equation}
\label{Regularity_Gauge_Field}
\iota_{\, V} \, A^I \rvert_{r = r_+} = 0 \, .
\end{equation}
This leads us to fix the constant gauge choice $\alpha^I$ introduced in \eqref{time_component_A} as 
\be\label{fix_gauge_5d}
\alpha^I = - \Phi^I\ ,
\ee
where $\Phi^I$ is the electrostatic potential \eqref{Entropy_Temperature_Potentials}. We describe the rest of the computation of the on-shell action in Appendix~\ref{app:OnShActionsAdS5} and  just provide the final result here:
\begin{align}
\label{On_Shell_Action}
I & = I_0 -\frac{\pi\beta}{12} \Big[ 2 m \left( c_1  s_1  \Phi^1 + c_2  s_2  \Phi^2 + c_3  s_3  \Phi^3 \right)  + 4 m^2  g ^2  \left(s_1^2 s_2^2 + s_1^2 s_3^2 + s_2^2 s_3^2 \right)  \notag \\[1mm]
&\qquad \quad \qquad   +   3 m  (g ^2  a^2 -1)  +   3 g ^2  r_+^4 + 2m\left(2g ^2  r_+^2  - 1  \right)\left(s_1^2 + s_2^2 + s_3^2 \right)  \Big]  \, ,
\end{align}
where 
\be\label{Casimir_Energy}
I_0 = \beta E_0
\ee
is the on-shell action of empty AdS$_5$ at temperature $\beta$.

One can check that the quantities above satisfy the quantum statistical relation:
\begin{equation}
\label{Quantum_Statistical_Relation}
I = \beta E -  S - \beta\,\Omega \, J - \beta\,\Phi^I \, Q_I  \, .
\end{equation}
From a microscopic point of view, this is interpreted as the relation between a grand-canonical partition function $I=-\log Z_{\rm grand}$, seen as a function of the chemical potentials, $I=I(\beta,\Omega,\Phi^I)$, and
the microcanonical partition function $S=\log Z_{\rm micro}$, seen as a function of the charges $S=S(E,J,Q_I)$. The charges are obtained by varying $I$ with respect to the chemical potentials as
\be
E = \frac{\partial I}{\partial\beta}\,,\qquad J = -\frac{1}{\beta}\frac{\partial I}{\partial \Omega}\,, \qquad Q_I = -\frac{1}{\beta}\frac{\partial I}{\partial \Phi^I}\,.
\ee

Let us comment on the contribution $E_0$ to the energy and the corresponding $I_0$ in the on-shell action. These are sensitive to the regularization adopted:
had we computed the action using the background subtraction method as done for similar solutions in  e.g.\ \cite{Hawking:1998kw,Gibbons:2004ai,Chen:2005zj}, we would have found the same result \eqref{On_Shell_Action}, but with $I_0=0$. Indeed background subtraction regularizes the divergence due to the infinite spacetime volume in a way different from holographic renormalization. It does so by subtracting the action of empty AdS space, with a boundary at large distance $\rnot$ matched to the boundary of the black hole solution, and then sends $\rnot\to \infty$. In this way the action $I$ is measured relative to the action of the AdS vacuum which results from taking $m=0$. Therefore in this approach $I_0=0$ by construction. 
The quantum statistical relation is still satisfied, provided one chooses $E_0=0$ for the AdS mass by the same logic.
Within the framework of holographic renormalization, one can shift $E_0$ (and $I_0$) to any desidered value by adding a finite, local counterterm to the action. Specifically, by adding to the Lorentzian action the finite boundary term $\frac{\varsigma}{8\pi}\int\diff^4x\sqrt{h} R^2$, where $\varsigma$ is a parameter, $h_{ij}$ is the boundary metric and $R$ its Ricci curvature, one obtains the shift $I_0 \to I_0 -9\pi g\beta\varsigma $ and $E_0 \to E_0-9\pi g\varsigma $.\footnote{This counterterm also yields a trivial $-\frac{3\varsigma}{2\pi}  \nabla^2 R$ contribution to the trace of the energy-momentum tensor, while the ``minimal subtraction'' scheme that we used to reach \eqref{eq:E0} is characterized by the fact that the trace of the holographic energy-momentum tensor does not contain trivial $\nabla^2 R$ terms.}
 In fact, the boundary field configuration of the solution we are considering implies that this is the only independent finite term that respects diffeomorphism and gauge invariance.
 Since it can be shifted by an arbitrary constant via a local counterterm, $E_0$ (and $I_0$) is an ambiguous quantity, which does not have an intrinsic meaning. The AdS/CFT correspondence identifies $E_0$ with the Casimir energy of the dual conformal field theory on $S^3\times \mathbb{R}$, and the same argument leads to conclude that this quantity does not have intrinsic value  \cite{Assel:2015nca}.
 However, the situation changes in the presence of additional symmetries, such as supersymmetry. In a supersymmetric setup the $R^2$ counterterm is not allowed, hence $E_0$ does acquire physical meaning \cite{Assel:2014tba,Assel:2015nca}. In fact, in \cite{Papadimitriou:2017kzw,An:2017ihs} $E_0$ has been interpreted as the consequence of a supercurrent anomaly, which is physical in nature. It may  be possible to see this  as a mixed anomaly and thus shift it away by adding local counterterms that restore supersymmetry at the expense of breaking part of the diffeomorphisms, along the lines of \cite{Genolini:2016sxe,Genolini:2016ecx,Closset:2019ucb}.\footnote{Similar considerations apply to the on-shell action discussed in \cite{Cabo-Bizet:2018ehj}, which was originally computed in \cite{Chen:2005zj} using the background subtraction method. We have checked that the same expression for the on-shell action is recovered if one uses holographic renormalization, again up to the $E_0$ vacuum energy factor.}

The solution admits an extremal limit. This can be seen by considering the function $Y(r)$ in the metric \eqref{Metric_CLP}. Being a cubic polynomial, this can be written as
\be
Y(r) = g^2(r^2 - r_+^2) (r^2 - r_0^2) (r^2 - r_-^2) \,,
\ee
where the roots  $r_+^2 \geq r_0^2 \geq r_-^2$ are related to the parameters of the solution as:
\begin{align}\label{rel_roots_params_5d}
r_+^2+r_0^2+r_-^2 &= -2m(s_1^2 + s_2^2 + s_3^2) -  g^{-2} \ ,\nn\\[1mm]
r_+^2r_0^2 + r_0^2r_-^2 + r_-^2r_+^2 &=   4m^2(s_1^2s_2^2 + s_2^2s_3^2+s_3^2s_1^2)+2m(a^2-g^{-2}) \ ,\nn\\[1mm]
r_+^2r_0^2r_-^2 &= -8m^3s_1^2s_2^2s_3^2 -g^{-2} f_3(r=0)    \ .
\end{align}
From \eqref{Entropy_Temperature_Potentials} we observe that the product of the temperature and the entropy is proportional to 
\be
T S \,=\, \frac{\pi}{8}\frac{Y'(r_+)}{r_+} \,=\, \frac{\pi g^2}{4}(r_+^2-r_0^2)(r_+^2-r_-^2)\,,
\ee 
hence the limit in which the roots $r_+^2$ and $r_0^2$ coalesce corresponds to the extremality condition $T=0$ (as long as the horizon area remains finite).
It is important to notice that this condition does not imply supersymmetry. We turn to supersymmetry next.

\subsection{The BPS solution}

It was found in \cite{Cvetic:2005zi} that one solution to the supergravity Killing spinor equations exists if the parameters satisfy:
\begin{equation}
\label{Sub_a_SUSY}
 a\,g \,=\, \frac{1}{\rme^{\delta_1+\delta_2+\delta_3}} \, .
\end{equation}
Hence the solution preserves two supercharges.
For simplicity, we set $g = 1$ from now on in this section (this can easily be restored by dimensional analysis).
We also find it convenient to trade the parameters $\delta_I$ for new parameters $\mu_I$, defined as
\begin{equation}
\label{delta_into_mu_map}
\rme^{4\delta_I} = \frac{\mu_I \left(\mu_J+2 \right) \left(\mu_K+2 \right)}{\left(\mu_I+2 \right) \mu_J \, \mu_K}   \,,
\end{equation} 
where the indices $I, J, K$ are never equal. 
In terms of the $\mu_I$, the supersymmetry condition \eqref{Sub_a_SUSY} reads
\begin{equation}
a = \left({\frac{\mu_1 \, \mu_2 \, \mu_3}{\left(\mu_1+2 \right) \left(\mu_2 + 2 \right) \left(\mu_3 + 2 \right)}} \right)^{1/4} \, .
\end{equation}
After this is imposed, closed timelike curves in the solution can be avoided by taking
\begin{equation}\label{m_BPS}
m = m_\star \equiv \frac{1}{2}\sqrt{\mu_1\mu_2\mu_3(\mu_1+2)(\mu_2+2)(\mu_3+2)} \, ,
\end{equation}
which using \eqref{rel_roots_params_5d} implies  that the horizons at $r_+$ and $r_0$ merge,
\be
r_0\to r_\star \leftarrow r_+\ ,
\ee
their common location being given by\footnote{We correct an overall minus sign typo in the corresponding expression given in eq.~(3.74) of~\cite{Cvetic:2005zi}.}
\be
r_\star^2 \equiv \frac{1}{2}\left(\sqrt{\mu_1\mu_2\mu_3(\mu_1+2)(\mu_2+2)(\mu_3+2)} - \mu_1\mu_2 - \mu_2\mu_3  - \mu_3\mu_1 -  \mu_1\mu_2\mu_3\right) \, .
\ee
Therefore the supersymmetry condition \eqref{Sub_a_SUSY} together with the requirement \eqref{m_BPS} of no causal pathologies implies extremality. We call the solution that is both supersymmetric and extremal the {\it BPS black hole} solution.

 The BPS solution thus obtained was first found in \cite{Gutowski:2004yv}, and depends on the three real parameters $\mu_I$, $I=1,2,3$. Regularity of the metric requires these to satisfy
\be\label{conditions_mu}
\mu_I > 0 \ ,\qquad 4\mu_1\mu_2\mu_3\,(\mu_1+\mu_2+\mu_3+1) > (\mu_1\mu_2 + \mu_2\mu_3 + \mu_3\mu_1)^2\ ,
\ee
which implies $r_\star^2>0$.
In our analysis below we will assume that the $\mu_I$ are chosen so that these conditions are satisfied.
The BPS value of the Bekenstein-Hawking entropy is~\cite{Gutowski:2004yv}:
\be\label{BPS_Entropy}
S^\star =  \frac{\pi ^2}{4} \, \sqrt{4\mu_1  \mu_2  \mu_3 \left(\mu_1 + \mu_2 + \mu_3 +1 \right)- \left(\mu_1  \mu_2 + \mu_2  \mu_3 + \mu_3  \mu_1 \right)^2} \, .
\ee
 In the BPS solution, the chemical potentials take the fixed values
\begin{equation}
\label{Chemical_Potential_BPS}
  \Omega^\star = 2 , \qquad  \Phi^{I\,\star} =  1 \, ,
\end{equation}
while the inverse temperature diverges, $\beta \to \infty$.
The BPS charges are
\begin{align}\label{BPS_Charges}
E^\star & = E_0 + \frac{\pi}{4} \,   \Big(2 \, \mu_1 \, \mu_2 \, \mu_3 + \frac{3}{2} \left(\mu_1 \, \mu_2 + \mu_2 \, \mu_3 + \mu_3 \, \mu_1 \right)+ \mu_1 + \mu_2 + \mu_3 \Big) \, , \notag \\[1mm]
J^\star & = \frac{\pi}{8}   \left(2 \, \mu_1 \, \mu_2 \, \mu_3 + \mu_1 \, \mu_2 + \mu_2\, \mu_3 + \mu_3 \,\mu_1 \right) \, , \notag \\[1mm]
Q_1^\star & = \frac{\pi}{8}  \left(2 \, \mu_1 + \mu_1\, \mu_2 + \mu_1\, \mu_3  - \mu_2 \, \mu_3 \right) \, ,
\end{align}
with $Q_2^\star$, $Q_3^\star$ being obtained from $Q_1^\star$ by a cyclic permutation of the indices $1,2,3$.
As a consequence of supersymmetry, the charges satisfy the linear relation
\begin{equation}
\label{Charge_BPS_Relation}
H^\star \equiv  E^\star - \Omega^\star \, J^\star - \Phi^{I\,\star} \, Q^\star_I = E_0  \, .
\end{equation}
In addition, the electric charges and angular momentum satisfy the non-linear relation 
\be\label{NonLinearConstraint}
Q_1^\star\, Q_2^\star\, Q_3^\star + \frac{\pi}{4}\, J^{\star \,2} = \left( Q_1^\star Q_2^\star+Q_2^\star Q_3^\star+Q_3^\star Q_1^\star - \frac{\pi}{2}J^\star \right)\left( Q_1^\star +Q_2^\star+ Q_3^\star  + \frac{\pi}{4} \right),
\ee
which is related to well-definiteness of the horizon area, that is of the entropy.
The BPS entropy can be written as a function of the charges in the suggestive form \cite{Kim:2006he}
\be\label{BPSentropy}
S^\star = 2\pi \sqrt{Q_1^\star Q_2^\star+Q_2^\star Q_3^\star+Q_3^\star Q_1^\star - \frac{\pi}{2} J^\star}\,.
\ee

\subsection{The BPS limit}

 Since in the BPS solution the chemical potentials take the fixed values $\Omega= \Omega^\star$, $\Phi^I=\Phi^{I\,\star}$, $\beta^{-1}=0$, one can ask if the BPS black hole satisfies non-trivial thermodynamic relations. In particular, one can ask what is the BPS version of the quantum statistical relation \eqref{Quantum_Statistical_Relation}.
In~\cite{Cabo-Bizet:2018ehj} it was shown how to define a BPS limit of black hole thermodynamics that reaches the BPS point along a supersymmetric trajectory in parameter space, and in this sense fully respects supersymmetry.  There are in fact many possible limits towards the BPS solution,  including the one previously proposed in \cite{Silva:2006xv}. However the limit of~\cite{Cabo-Bizet:2018ehj}, that respects supersymmetry all along the trajectory approaching the BPS locus, yields a result that agrees with dual supersymmetric field theory computations. We thus eliminate the parameter $a$ by imposing the supersymmetry condition \eqref{Sub_a_SUSY} and for the moment do not demand \eqref{m_BPS}, which as we reviewed above would imply extremality. Although for $m\neq m^\star$ the Lorentzian solution has closed timelike curves, we are interested in saddle points of the quantum gravity path integral, and thus allow ourselves to work with a complex section of the solution, to be specified momentarily. The solution now depends on the four parameters $m, \, \mu_1, \, \mu_2, \, \mu_3$. We wish to trade $m$ for the outer horizon position $r_+$ by  solving the equation $Y(r_+)=0$ for $m$; this will make it easier to study the limit towards extremality.  
 Since the equation $Y(r_+)=0$ is of third order in $m$, its solution is quite complicated. To circumvent this complication, we first change the radial coordinate $r$ into a new coordinate $R$, such that:\footnote{In terms of the old parameters $\delta_I$, the change of coordinate is expressed as $r^2  = R^2 - 2m \sinh^2\delta_1.$ This implies $r^2H_1= R^2$. The new coordinate $R$ should not be confused with the Ricci scalar.}
\begin{equation}\label{Change_r_to_R}
r^2  = R^2 + \frac{m}{m_\star} \left(r_\star^2 - \mu_1 \right)  \, .
\end{equation}
Although this breaks the symmetry in $\mu_1,\mu_2,\mu_3$, the latter will be restored in the final results after reaching the BPS point. 
The position of the outer horizon is now given by the largest root $R_+$ of the equation $Y(R) = 0$.
From \eqref{Change_r_to_R} we see that in the new coordinate the BPS horizon is found at
\be
R_\star^2 = \mu_1\ .
\ee
Now the equation $Y(R_+)=0$ is only quadratic in $m$, and its solution can be written as:
\begin{equation}
\label{Solution_m}
m = \frac{2 m_\star \, R_+^4  (R_+^2 + 1 ) }{R_+^4 \left(2  \mu_1 - \mu_2 - \mu_3 \right) + R_+^2 \left( \mu_1 \mu_2 + \mu_2  \mu_3 + \mu_3  \mu_1 + 2 \mu_1 \right)-\mu_1  \mu_2  \mu_3\mp\! (  R_+^2 - \mu_1 )  \CR } ,
\end{equation}
where we introduced the quantity:
\begin{equation}
\label{Square_root}
\CR = \sqrt{  R_+^4 (\mu_2 - \mu_3)^2 - 2 \,R_+^2 \mu_2 \, \mu_3  \left(\mu_2 + \mu_3+2 \right)+ \mu_2^2 \, \mu_3^2}\ .
\end{equation}
 Due to the undefined sign of the argument of this square root, the expression for $m$ in \eqref{Solution_m} may be complex. For very large $R_+^2$ we have that $\CR$ is real, thus $m$ is real. On the other hand, for $R_+^2$ sufficiently close to $R_\star^2 =\mu_1$, that is sufficiently close to the extremal value,  the square root $\CR$ is purely imaginary as a consequence of \eqref{conditions_mu}. Therefore close to  extremality $m$ takes a complex value. In the strict extremal limit $R_+^2 = R_\star^2$ we have that the factor multiplying $\CR$ in \eqref{Solution_m} goes to zero, so although $\CR$ is purely immaginary, $m$ becomes real and reaches its BPS value \eqref{m_BPS}.  

Fixing $a$ as in~\eqref{Sub_a_SUSY} and trading  $m$ for $R_+$ as in~\eqref{Solution_m} identifies our family of complexified, supersymmetric solutions. Evaluating the quantities~\eqref{Entropy_Temperature_Potentials} in this family of solutions we obtain quite cumbersome expressions, that we will not display here.  
Remarkably, we find that the chemical potentials obtained in this way satisfy the constraint:
\begin{equation}
\label{Constraint_Chemical}
\beta\, (1 + \Omega - \Phi^1 - \Phi^2 - \Phi^3 ) = \mp \, 2 \, \pi \, i \, ,
\end{equation}
where the sign choice follows from the one in \eqref{Solution_m}. 
Although we did not manage to derive \eqref{Constraint_Chemical} in full generality due to the complexity of the expressions for the chemical potentials, we verified it with many numerical checks over a wide range of the parameters as well as in a perturbative expansion near the BPS point. We find that eq.~\eqref{Constraint_Chemical} is satisfied in two slightly different ways, depending on the value of $R_+^2$. As we described above, for sufficiently large $R_+^2$, $m$ is real; however in this case $\beta$  is purely imaginary, so that \eqref{Constraint_Chemical} holds true.\footnote{We recall that $\beta$ is given by the expression in \eqref{Entropy_Temperature_Potentials}, so it is purely imaginary when $f_1$ is negative. We find that this happens precisely in the regime where $m$ is real. In this discussion we are assuming that $R_+^2$ is real.} On the other hand, when $R_+^2$ is close to the extremal value, $m$ is complex, and so are the chemical potentials $\beta,\Omega,\Phi^I$; still \eqref{Constraint_Chemical}  is satisfied. 

In order to make the near-extremal behavior of the different quantities appearing in \eqref{Constraint_Chemical} more explicit, we set $R_+ = R_\star + \epsilon$ and study the limit $\epsilon\to 0$. We perform the computation choosing the upper sign in \eqref{Solution_m} for simplicity.
We find that in this limit the ``inverse temperature'' $\beta$ diverges as: 
\begin{equation}
\label{Beta_around_BPS}
\beta = \frac{4 \, S^\star + i \, \pi^2  \, \left[2 \, \mu_1^2 - \mu_2 \, \mu_3 + \mu_1 \left(2 + \mu_2 + \mu_3 \right) \right]}{4 \, \pi \, \epsilon\, \sqrt{\mu_1} \left(1 + \mu_1 + \mu_2 +  \mu_3 \right)} \, + \, \CO(\epsilon^0) \, ,
\end{equation}
where $S^\star$ is the BPS entropy given in \eqref{BPS_Entropy}.
Hence $\beta$ is complex at leading order near the BPS point.  
The same holds for the other chemical potentials, which read
\begin{align}
\Omega & = \Omega^\star -\frac{   4 \, S^\star - i \, \pi^2 \left( \mu_1  \mu_2 + \mu_3  \mu_1 -\mu_2  \mu_3  \right) }{S^\star \, \sqrt{\mu_1} \, \left(1+\mu_1  \right) }\,\epsilon + \CO(\epsilon^2) \, , \notag \\
\Phi^1 & = \Phi^{1 \, \star} + \frac{ 4  S^\star \left[\mu_1 \mu_2 + \mu_3\mu_1  - \mu_2  \mu_3 \right] - 2i\pi^2 \left[ (1+\mu_1)\mu_1\mu_2\mu_3  -8(S^{\star}/\pi^{2})^2 \right]}{2 \, S^\star \, \mu_1^{5/2} \left( 1+\mu_1  \right) \mu_2 \, \mu_3  }\,\epsilon + \CO(\epsilon^2), 
\end{align}
with $\Phi^2,\Phi^3$ being obtained from $\Phi^1$ by a cyclic permutation of the $\mu_I$. It follows that 
\begin{equation}
\label{Potentials_around_BPS}
1 + \Omega - \Phi^1 - \Phi^2 - \Phi^3 = \frac{- 4 \, \mu_1^2 + 2 \, \mu_2 \, \mu_3 - 2 \, \mu_1 \left(2 + \mu_2 + \mu_3 \right) - 8 \, i  \, S^\star/\pi^2  }{\mu_1^{3/2} \, \left(1 + \mu_1 \right) } \, \epsilon \, + \, \CO(\epsilon^2) \, .
\end{equation}
Multiplying the complex quantities~\eqref{Beta_around_BPS} and~\eqref{Potentials_around_BPS}, the factors of $\epsilon$ cancel out and one can easily check that the finite result~\eqref{Constraint_Chemical} is obtained. 

Eq.~\eqref{Constraint_Chemical} was understood in~\cite{Cabo-Bizet:2018ehj} as a regularity condition for the Killing spinor of the supersymmetric solution, ensuring that this is {\it antiperiodic} around the Euclidean time circle of finite length $\beta$ corresponding to the orbit of the Killing generator $V$ of the horizon given in \eqref{Killing_Vector}, when the regular gauge $\iota_V A^I|_{r_+}=0$ is assumed. In fact the only spin structure allowed in the topology of the cigar formed by the radial direction and the orbit of $V$, which shrinks to zero size as $R\to R_+$, is the one of an antiperiodic spinor.

We now introduce new chemical potentials $\omega$ and $\varphi^I$ by redefining the previous ones as:
\begin{equation}
\label{Little_Potentials}
\omega = \beta \left(\Omega - \Omega^\star \right) , \qquad \varphi^I = \beta \left(\Phi^I - \Phi^{I\,\star} \right) \, .
\end{equation}
We also introduce the supersymmetric Hamiltonian 
\begin{align}
H  &= E - \Omega^\star J - \Phi^{I\,\star}\,Q_I \nn\\[1mm]
&= E - 2J - Q_1-Q_2-Q_3\ .
\end{align}
While $E$ is the charge for translations generated by  $\frac{\partial}{\partial t}$, the supersymmetric Hamiltonian $H$ is the charge for translations generated by the Killing vector $K= 
\frac{\partial}{\partial t}+ \Omega^\star \frac{\partial}{\partial \psi}$ that arises as a bilinear of the Killing spinor, covariantized by the term $\iota_K A^I|_\infty Q_I = - \Phi^{I\,\star}\,Q_I$.
This is related to the anticommutator of the supercharges by $\{\mathcal{Q},\overline{\mathcal{Q}}\}= H - E_0$, where $E_0$ is the anomalous term induced by the supercurrent anomaly (in a renormalization scheme that preserves diffeomorphism and gauge invariance) \cite{Papadimitriou:2017kzw,An:2017ihs}.
Using the new variables, the quantum statistical relation~\eqref{Quantum_Statistical_Relation} can be expressed as:
\begin{align}
\label{QSR_manipulating}
I &  = \beta  H - S -  \omega\, J -  \varphi^I \, Q_I \, .
\end{align}
We then see that $\omega$ and $\varphi^I$ are the chemical potentials conjugate to $J$ and $Q_I$, respectively, when the time translations are generated by the supersymmetric Hamiltonian $H$.
Since in any supersymmetric solution $\{\mathcal{Q},\overline{\mathcal{Q}}\}$ evaluates to zero and thus $H=E_0$, we arrive at the \emph{supersymmetric quantum statistical relation}
\begin{equation}
\label{QSR_SUSY}
I - I_0 = - S -  \omega\, J -  \varphi^I \, Q_I 	\, ,
\end{equation}
where we used $I_0=\beta E_0$. The constraint~\eqref{Constraint_Chemical} now reads:
\begin{equation}
\label{Constraint_Little}
\omega - \varphi^1 - \varphi^2 - \varphi^3 = \mp \, 2\, \pi \, i  \, .
\end{equation}
We observe that varying the supersymmetry relation between the charges and subtracting this from the first law \eqref{first_law_5d}, we obtain a supersymmetric form of the first law:
\begin{equation}
 \diff S + \omega \, \diff J + \varphi^I \, \diff Q_I =0 \, .
\end{equation}
Moreover, plugging the supersymmetric condition~\eqref{Sub_a_SUSY} and the expression~\eqref{Solution_m} for $m$ into the on-shell action~\eqref{On_Shell_Action}, we find that the latter takes the simple form:
\begin{equation}
\label{On_Shell_Action_With_Potentials}
I - I_0 = \pi \,\frac{ \varphi^1  \varphi^2  \varphi^3}{(\omega)^2} \, .
\end{equation}
Notice that the right hand side of~\eqref{On_Shell_Action_With_Potentials} is independent of $\beta$.

The supersymmetric charges $E$, $J$ and $Q_I$ are evaluated by substituting the supersymmetry condition~\eqref{Sub_a_SUSY} and the formula \eqref{Solution_m} for  $m$ in \eqref{nonextremal_charges}. We checked that although the expressions thus obtained are generically complex, they satisfy the supersymmetry relation $H=E_0$. In the same way we obtain a generically complex expression for the entropy. The fact that the entropy (that is the area of the horizon) is complex is related to the fact that when continued back to Lorentzian signature, the supersymmetric but non-extremal solution presents a pseudo-horizon rather than a horizon \cite{Cvetic:2005zi}.

We can now take the limit to extremality by sending $R_+ \to R_\star$. Doing this, our complexified family of supersymmetric solutions reaches the real, BPS  solution of \cite{Gutowski:2004yv}. All the main physical quantities become real; in particular the entropy and the charges take the values \eqref{BPS_Entropy}, \eqref{BPS_Charges}.

In the extremal limit $R_+\to R_\star$ the temperature vanishes, therefore $\beta$ diverges; at the same time, $\Omega\to \Omega^\star$, $\Phi^I\to \Phi^{I\,\star}$, in such a way that the chemical potentials $\omega$, $\varphi^I$ defined in \eqref{Little_Potentials} stay finite. We denote the BPS values of the redefined chemical potentials as
\begin{equation}
\label{BPS_Little_Potentials_Def}
\omega^\star = \lim_{R_+ \to R_\star} \omega \, , \qquad \varphi^{I\,\star} = \lim_{R_+ \to R_\star} \varphi^{I} \, .
\end{equation}
By evaluating these limits we obtain
\begin{align}
\label{BPS_Little_Potentials}
\omega^\star & = \frac{-2\pi}{\mu_1 + \mu_2 + \mu_3 +1 }\bigg[ \frac{ \mu_1 \, \mu_2 + \mu_2\, \mu_3 + \mu_3 \, \mu_1 }{  \sqrt{ 4 \mu_1 \mu_2 \mu_3 \left(\mu_1+\mu_2 + \mu_3 + 1 \right) - (\mu_1\mu_2 +\mu_2 \mu_3 + \mu_3\mu_1)^2  }} \pm  i \bigg]  , \notag \\[3mm] 
\varphi^{1\,\star} & = \frac{\pi}{\mu_1+ \mu_2 + \mu_3 + 1 }\bigg[\frac{  \mu_1 ( \mu_2^2 + \mu_3^2) - \mu_2 \, \mu_3(\mu_2+\mu_3+2)  }{\sqrt{ 4 \mu_1 \mu_2 \mu_3 \left(\mu_1+\mu_2 + \mu_3 + 1 \right) - (\mu_1\mu_2 +\mu_2 \mu_3 + \mu_3\mu_1)^2  }} \nn\\[1mm] 
&\qquad \qquad \qquad \qquad \quad \pm  i  \left(\mu_2+ \mu_3 \right)  \bigg] \, ,
\end{align}
with the expressions for $\varphi^{2\,\star}$ and $\varphi^{3\,\star}$ being obtained from the one for $\varphi^{1\,\star}$ through straightforward permutations of the indices $1,2,3$.
Therefore these chemical potentials remain complex even after the BPS limit is taken.\footnote{Note that the argument of the square roots in \eqref{BPS_Little_Potentials} is positive due to assumption \eqref{conditions_mu}, and proportional to the BPS entropy \eqref{BPS_Entropy}.} We remark that  $\omega$ and $\varphi^I$ are not the leading order terms~\eqref{Chemical_Potential_BPS} of the chemical potentials $\Omega$ and $\Phi^I$ in the BPS limit. They are instead the next-to-leading-order terms in the expansion of $\Omega$ and $\Phi^I$ around their BPS value:
\be
\label{Little_Potential_Expansion}
\Omega = \Omega^\star + \frac{1}{\beta}\, \omega^\star + \dots \, , \quad \qquad
\Phi^I = \Phi^{I\,\star} + \frac{1}{\beta}\, \varphi^{I\,\star} + \dots \, .
\ee
Since the limit is smooth, these BPS chemical potentials still satisfy the constraint
\begin{equation}
\label{Constraint_Little_BPS}
\omega^\star - \varphi^{1\,\star} - \varphi^{2\,\star} - \varphi^{3\,\star} = \mp \, 2 \, \pi\, i \, ,
\end{equation}
and the on-shell action of the BPS solution reads
\begin{equation}
\label{On_Shell_Action_With_Potentials_BPS}
(I  - I_0)^\star= \pi \,\frac{ \varphi^{1\,\star}  \varphi^{2\,\star}  \varphi^{3\,\star}}{(\omega^\star)^{\, 2}} \, .
\end{equation}

As argued in~\cite{Cabo-Bizet:2018ehj}, the supersymmetric on-shell action $I-I_0$, seen as a function of the chemical potentials \eqref{Little_Potentials}, should be regarded as minus the logarithm of the supersymmetric grand-canonical partition function in the semi-classical approximation to the quantum gravity path integral. Therefore, its Legendre transform must be the logarithm of the microcanonical partition function, that is the entropy. This gives a physical derivation of the extremization principle proposed in~\cite{Hosseini:2017mds}. 

The Legendre transformation is not completely straightforward because of the constraint~\eqref{Constraint_Little} between the chemical potentials. It was described in detail in~\cite[Appendix B]{Cabo-Bizet:2018ehj}\footnote{The variables used in~\cite[Appendix B]{Cabo-Bizet:2018ehj} are related to the ones used here as:
\begin{equation}
\omega_1 = \omega_2 = \frac{\omega}{2} \, , \qquad \Delta_I = - \Phi_I \, , \qquad \mu = - \frac{\pi}{4} \, , \qquad n = \mp 1 \, , \qquad J_1 = J_2 = J \, ,  \qquad
Q_I^\text{there} = - Q_I^{\rm here} \,. \nn
\end{equation}} 
and we recall here the main steps. The supersymmetric quantum statistical relation~\eqref{QSR_SUSY} can be written as
\begin{equation}
\label{QSR_BPS_Lagrange}
I - I_0 = - S -  \omega \, J - \varphi^{I} \, Q_I - \Lambda \left(\omega - \varphi^{1} - \varphi^{2} - \varphi^{3} \pm \, 2  \, \pi\, i \, \right) \, ,
\end{equation}
where the constraint~\eqref{Constraint_Little} is enforced through a Lagrange multiplier $\Lambda$. Although the constraint is identically satisfied in the supersymmetric solution, at this stage we are not assuming any explicit expression for $\omega,\varphi^I$, since we want to treat them as the basic variables to be varied. 
 Extremizing \eqref{QSR_BPS_Lagrange} with respect to $\Lambda,\omega,\varphi^I$ we retrieve the constraint~\eqref{Constraint_Little}, together with the equations
\begin{equation}
\label{Conjugated_BPS}
- \frac{\partial (I - I_0)}{\partial \omega} = J + \Lambda \, , \qquad - \frac{\partial (I - I_0)}{\partial \varphi^{I}} = Q_I + \Lambda \, , \quad I=1,2,3\,,
\end{equation}
which state the conjugacy relation between supersymmetric charges and chemical potentials. These five equations can be solved for $\omega,\varphi^I$ and $\Lambda$ in terms of the charges $J,Q_I$ (see \cite{Cabo-Bizet:2018ehj} for the explicit expressions). By substituting the solution in \eqref{QSR_BPS_Lagrange}, one obtains a formula for the entropy $S$ that is a complex function of the charges. Further demanding reality of $S$, as well as of $J,Q_I$, one obtains precisely the non-linear relation \eqref{NonLinearConstraint} between the BPS charges, together with the expression \eqref{BPSentropy} for the Bekenstein-Hawking entropy of the BPS black hole. The reality condition on $S$ may be understood as a well-definiteness condition for the horizon area, and this is what leads to \eqref{NonLinearConstraint} in the extremization procedure. 

We have verified that this extremization is realized in the black hole solution. In particular, we checked that our BPS chemical potentials \eqref{BPS_Little_Potentials} match the saddle point value of $\omega,\varphi^I$ obtained by solving the extremization equations \eqref{Conjugated_BPS} in terms of the charges $J,Q_I$, demanding reality of the entropy, and substituting the parameterization \eqref{BPS_Charges} of the BPS charges. We also checked that this match still holds true when one compares the supersymmetric but non-extremal values of $\omega,\varphi^I$, even if in this case the entropy and the charges are generically complex.

\section{Rotating, electrically charged AdS$_4$ black holes}\label{4d_section}

In this section we consider a class of rotating, electrically charged, asymptotically AdS$_4$ black holes with an uplift to M-theory. We show that a limiting procedure analogous to the one discussed in five dimensions leads to complexified chemical potentials satisfying the constraint \eqref{constraint_intro} and to an on-shell action that matches the entropy function recently proposed in~\cite{Choi:2018fdc}. 

\subsection{The non-supersymmetric, finite temperature solution}\label{sec:nonsusysol_4d}

The black hole solution of interest was first constructed in \cite{Chong:2004na} within a consistent truncation of four-dimensional $\mathcal{N}=8$, $\SO(8)$ gauged supergravity. The truncation is obtained by restricting to the $\U(1)^4$ Cartan subgroup of $\SO(8)$ and setting the corresponding four gauge fields pairwise equal. We start with a brief review of the solution, again referring to the presentation of \cite{Cvetic:2005zi}. 
The action is:
\begin{align}
\label{Lagrangian_Four_Dim}
\CS & = \frac{1}{16\pi}\int \Big[\left(R - 2 \, \CV  \right)  \star  1 \, - \frac{1}{2} \, \diff \scal \wedge \star\, \diff \scal - \frac{1}{2} \, \rme^{2 \scal} \, \diff \chi \wedge \star \,\diff \chi  - \frac{1}{2} \, \rme^{-\scal} \, F_3 \wedge \star F_3  \notag \\
& \qquad\qquad    - \frac{1}{2} \, \chi \, F_3 \wedge F_3  - \frac{1}{2 \left(1 + \chi^2 \rme^{2 \scal} \right)} \left( \rme^\scal \,  F_{1} \wedge \star F_1 - \rme^{2 \scal} \, \chi \, F_1 \wedge F_1 \right) \Big] \, ,
\end{align}
where $F_1$ and $F_3$ are the field strengths of the Abelian gauge fields $A_1=A_2$ and $A_3=A_4$,\footnote{In this section we use lower indices on the vector fields and the respective chemical \hbox{potentials $\Phi$.}} and $\CV$ is the scalar potential for the axion and dilaton scalar fields $\chi,\xi$:
\begin{equation}
\label{Scalar_Potential_Four_Dim}
\CV = - \frac{1}{2} \,g^2 \left(4 + 2 \cosh{\scal} + \rme^\scal \chi^2 \right) \, .
\end{equation}
The solution uses coordinates $(t,r,\theta,\phi)$, where $\theta\in [0,\pi]$, $\phi\sim\phi+2\pi$ parameterize a two-sphere. In a frame that rotates at infinity, the metric reads:
\begin{equation}
\label{Metric_Four_Dimensional}
\diff s_4^2 = - \frac{\Delta_r}{W} \left( \diff t - \frac{a}{\Xi}  \sin^2{\theta} \, \diff \phi \right)^2 + W \Big( \frac{\diff r^2}{\Delta_r} + \frac{\diff \theta^2}{\Delta_\theta}  \Big) + \frac{\Delta_\theta  \sin^2 \theta}{W} \Big( a \, \diff t - \frac{ r_1 r_2 + a^2 }{\Xi}  \diff \phi \Big)^2 , 
\end{equation}
where
\begin{align}
\label{Quantites_Four_Dimensional}
& r_i = r + 2 \, m \, s_i^2 \, , \notag \\
& \Delta_r = r^2 + a^2 - 2 \, m \, r + g^2 \, r_1 \, r_2 \left(r_1 \, r_2 + a^2 \right) \, , \notag \\
& \Delta_\theta = 1 -  a^2 g^2\, \cos^2 \theta \, , \qquad W = r_1 \, r_2 + a^2 \, \cos^2 \theta \, , \qquad \Xi = 1 - a^2  g^2 \, ,
\end{align}
and $s_i = \sinh \delta_i$, $c_i = \cosh \delta_i$, $i=1,2$.
The scalar fields are given by
\begin{equation}
\label{Scalar_Phi_Chi}
\rme^{\scal} = 1 + \frac{r_1 \left(r_1 - r_2 \right) }{W} \, , \qquad \chi = \frac{a \left(r_2 - r_1 \right) \cos \theta }{r_1^2 + a^2 \cos^2 \theta } \, ,
\end{equation}
while the gauge fields read
\begin{equation}
\label{Gauge_Fields_4_dim}
A_1 = \frac{2 \sqrt{2} \, m \, s_1  c_1  r_2  }{W}\Big( \diff t - \frac{a}{\Xi} \, \sin^2 \theta  \, \diff \phi \Big) \, , \quad A_3 = \frac{2 \sqrt{2} \, m \, s_2  c_2  r_1 }{W}\Big( \diff t - \frac{a}{\Xi} \, \sin^2 \theta \,  \diff \phi \Big)  \, .
\end{equation}
The solution is thus controlled by four parameters $m,a,\delta_1,\delta_2$.
Since it is contained in $\SO(8)$ gauged supergravity, the solution uplifts to eleven-dimensional supergravity on $S^7$ (see \cite{Chong:2004na} and references therein for the explicit uplift formulae). The dual SCFT$_3$ is then the ABJM theory.
 
The solution has an outer horizon at $r=r_+$, defined as the largest root of $\Delta_r$. This is a Killing horizon, generated by the vector 
\be
V = \frac{\partial}{\partial t'} + \Omega \, \frac{\partial}{\partial \phi'} \,,
\ee 
where the coordinates
\be\label{Change_To_Non_Rotating}
\phi' = \phi + a \,g^2\, t\,,\qquad t'=t 
\ee 
define a frame that is non-rotating at infinity, and
\be\label{AngularV_4d}
\Omega =  \frac{a \left(1 + g^2 \, r_1 \, r_2 \right) }{r_1 \, r_2 + a^2}
\ee
is the angular velocity of the horizon. The Bekenstein-Hawking entropy, the  inverse temperature  and the electrostatic potentials of the black hole are given by:
\begin{align} \label{Properties_Solution_Four_Dim}
&\ \ S = \frac{ \pi \left(r_1 \, r_2 + a^2 \right)}{\Xi} \, ,  \qquad \beta = 4 \pi \left(r_1 \, r_2 + a^2\right)  \left(\frac{\diff \Delta_r}{\diff r} \right)^{-1}\, , \notag\\[1mm]
\Phi_1 &= \Phi_2 = \frac{2 \, m \, s_1 \, c_1 \, r_2}{r_1 \, r_2 + a^2} \, , \qquad  \Phi_3 = \Phi_4 = \frac{2  \, m  \,  s_2 \,  c_2  \, r_1}{r_1 \, r_2 + a^2} \, ,
\end{align}
where all the functions of the radial coordinate are evaluated in $r_+$. Here the electrostatic potentials $\Phi_I$, $I=1,\ldots,4$, are obtained from the four vector fields gauging the Cartan subgroup of $\SO(8)$. Since these are set pairwise equal in the action \eqref{Lagrangian_Four_Dim},  necessarily  we have $\Phi_1 = \Phi_2$ and $\Phi_3 = \Phi_4$. 
The energy (that is the charge associated with translations generated by $\frac{\partial}{\partial t'}$), the angular momentum (that is the charge associated with rotations generated by $-\frac{\partial}{\partial \phi'}$) and the electric charges are:
\begin{align}
\label{Energy, Angular Momentum, Charges_Four_Dim}
& E = \frac{m}{\Xi^2} \left(1 + s_1^2 + s_2^2 \right) \, ,\qquad
 J = \frac{m \, a}{\Xi^2} \left(1 + s_1^2 + s_2^2 \right) \, , \notag \\[2mm]
&\ \quad Q_1 = Q_2 = \frac{m \, s_1 \, c_1 }{2 \, \Xi}  \, , \qquad
 Q_3 = Q_4 = \frac{m \, s_2 \, c_2}{2 \, \Xi}  \, . 
\end{align}
The electric charges and the angular momentum were obtained in \cite{Cvetic:2005zi} evaluating the standard Maxwell and Komar asymptotic integrals, respectively, while the energy was computed by integrating the first law of thermodynamics,
\begin{equation}\label{first_law_4d}
\diff E = T \, \diff S + \Omega \, \diff J + 2 \, \Phi_1 \, \diff Q_1 +2\,\Phi_3 \, \diff Q_3 
 \, .
\end{equation}
In  Appendix~\ref{app:OnShActionsAdS4} we check that the same expressions for the charges are obtained using holographic renormalization. We also compute the Euclidean on-shell action $I$ by the same method and find the result:
\begin{align}
\label{On_Shell_Action_Euclidean_Four_Dim}
I & = \frac{\beta}{2(a^2g^2-1)}\, \bigg\{  g^2r_+^3 + 3 \, m \,g^2 r_+^2 \left(s_1^2 + s_2^2 \right) + r_+ \left[a^2g^2 + 2 \, m^2g^2 \left(s_1^4 + 4  s_1^2  s_2^2 + s_2^4 \right) \right]  \nn\\[1mm]
&\quad \qquad\qquad\qquad  + m  \left(a^2g^2 + 4 \, m^2g^2 s_1^2 s_2^2 -1 \right) (s_1^2+s_2^2)  - m   \notag \\[1mm]
&\quad\qquad\qquad\qquad + \frac{2\,m^2 \left[ c_1^2  s_1^2 \left(2 \, m \, s_2^2 + r_+ \right) + c_2^2  s_2^2 \left(2 \, m \, s_1^2 + r_+ \right)\right]}{ a^2+\left(2 \, m \, s_1^2 + r_+ \right) \left(2 \, m \, s_2^2 + r_+ \right)} \,  \bigg\}  \, .
\end{align}
We have explicitly verified that the on-shell action and the quantities \eqref{AngularV_4d}, \eqref{Properties_Solution_Four_Dim}, \eqref{Energy, Angular Momentum, Charges_Four_Dim} satisfy the quantum statistical relation
\begin{equation}
\label{Quantum_Statistical_Relation_Four_Dimensional}
I = \beta E  -  S - \beta\, \Omega \, J - 2\beta \,  \Phi_1 \, Q_1 - 2\beta  \, \Phi_3 \,  Q_3  \, .
\end{equation}

\subsection{The BPS solution}

We will set $g=1$ from now on.
The solution presented above is supersymmetric if\footnote{In \eqref{susy_cond_4d} and \eqref{mstar_4d}, we are using the expressions given in  \cite{Chow:2013gba}, which correct typos in the corresponding expressions of \cite{Cvetic:2005zi}.}
\begin{equation}\label{susy_cond_4d}
a = \frac{2}{\rme^{2 \, \left(\delta_1 + \delta_2 \right)} - 1} \, .
\end{equation}
In the following we assume this condition and use it to eliminate $a$ from all expressions. We thus have a supersymmetric family of solutions described by the remaining parameters $m,\delta_1,\delta_2$. It was shown in \cite{Cvetic:2005zi} that for real values of these parameters, the equation $\Delta_r(r)=0$ determining the existence of a horizon only has a solution~if
\begin{equation}\label{mstar_4d}
m^2 = m_\star^2 \equiv \frac{\cosh^2 (\delta_1 + \delta_2 )}{4\, \rme^{\delta_1 + \delta_2} \sinh^3(\delta_1+\delta_2) \, c_1 \, s_1 \, c_2 \,  s_2  } \, ,
\end{equation}
in which case there is a regular horizon at
\begin{equation}\label{rstar_4d}
r=r_\star \equiv \frac{2 \, m_\star \, s_1 \, s_2}{\cosh(\delta_1+\delta_2)} \, .
\end{equation}
This is a double root of $\Delta_r$, hence the supersymmetric solution becomes  extremal and the temperature vanishes. This gives a BPS solution that is regular on and outside the horizon.\footnote{Specializing to $\delta_1=\delta_2$ gives a solution to pure gauged $\mathcal{N}=2$ supergravity in four dimensions, originally discussed in \cite{Kostelecky:1995ei,Caldarelli:1998hg}. We also remark that the BPS black holes we are discussing are different from the rotating solutions with magnetic charge recently found in \cite{Hristov:2018spe}.}

The chemical potentials take the BPS values
\begin{equation}
\label{Chemical_Potential_BPS_Four_Dimensions}
\Omega^\star = 1 \,, \qquad  \Phi_1^\star = \Phi_3^\star =  1 \, ,\qquad  \beta \to \infty  \, ,
\end{equation}
while the BPS charges are:
\begin{align}\label{BPS_charges_4d}
E^\star &= \frac{\left(c_1 \, c_2 - s_1 \, s_2 \right) \sqrt{\rme^{-(\delta_1 + \delta_2)} \left(c_1 \,  s_2 + c_2 \,  s_1 \right) } } {2 \left(\coth \left( \delta_1 + \delta_2 \right)-2 \right)^2 \sqrt{c_1 \, c_2  \, s_1 \, s_2}} \, ,\nn\\[1mm]
J^\star &= \frac{c_1 \, c_2 - s_1 \, s_2} {2 \left(\coth (\delta_1 + \delta_2) - 2 \right)^2 \sqrt{\rme^{3 \left(\delta_1 + \delta_2 \right)} \, c_1 \,  c_2  \, s_1 \, s_2   \left(c_1 \, s_2 + c_2 \, s_1 \right) }} \, ,\nn\\[1mm]
Q_1^{\star} &= \frac{\sqrt{c_1 \, c_2 \, s_1 \, s_2  \left(\rme^{2 \left(\delta_1 + \delta_2 \right)} - 1 \right) } } {2 \, \sqrt{2} \, c_2 \, s_2 \left(\rme^{2 \left(\delta_1 + \delta_2 \right)}-3\right)} \, ,
\end{align}
with $Q_3^\star$ being obtained from $Q_1^\star$ by switching $s_1 \leftrightarrow s_2$ and $c_1 \leftrightarrow c_2$. 
These satisfy the relation
\begin{equation}
\label{Charge_BPS_Relation_Four_Dimensions}
E^\star - \Omega^\star  J^\star - 2 \,\Phi^\star_1 \, Q^\star_1 -2\, \Phi^\star_3 \, Q^\star_3  = 0  \, ,
\end{equation}
that is a consequence of the supersymmetry algebra.
The BPS angular momentum and electric charges also satisfy \cite{Choi:2018fdc}
\begin{equation}\label{nonlinear_charges_4d}
J^\star = (Q_1^\star+Q_3^\star)\big(\sqrt{1+64 Q_1^\star Q_3^\star}-1\big)\,,
\end{equation}
which are related to the fact the we have imposed \eqref{mstar_4d} on top of the supersymmetry condition \eqref{susy_cond_4d} (having fixed two of the four free parameters of the solution, there cannot be more than two independent charges).
The BPS entropy reads
\begin{equation}
S^\star = \frac{2 \, \pi} {\rme^{2\delta_1 + 2\delta_2 } -3} \, ,
\end{equation}
and can  be expressed in terms of the charges as \cite{Choi:2018fdc}:
\begin{equation}
S^\star =  \frac{\pi\,J^\star}{2(Q_1^\star + Q_3^\star)} \,.
\end{equation}
Note that positivity of the BPS entropy restricts the allowed range of $\delta_1+\delta_2$.

\subsection{The BPS limit}

Our BPS limit proceeds similarly to the five-dimensional case.
We start by imposing the supersymmetry condition \eqref{susy_cond_4d}, while for the moment we do not require  \eqref{mstar_4d}. The equation $\Delta_r(r)=0$, with $\Delta_r$  being given in \eqref{Quantites_Four_Dimensional}, can be solved in a more general way than \eqref{mstar_4d}, \eqref{rstar_4d} if we allow for complex values of the parameter $m$.
In fact,  $\Delta_r(r)=0$ can be seen as an equation for $m$, where the solution depends on $\delta_1,\delta_2$ and on the position of the outer horizon, $r_+$. Since $\Delta_r(r)$ is a quartic polynomial in $m$, the solutions are rather cumbersome. The analysis is simplified if we change the radial coordinate $r$ into a new coordinate $R$, defined as:
\begin{equation}
\label{Change_r_to_R_Four_Dimensions}
r = R - 2 \, m \, s_1^2 \, .
\end{equation}
The equation for the horizon becomes $\Delta_r(R)=0$, and we denote by $R_+$ the position of the outer horizon.  This equation is now only quadratic in $m$, and its solution is:
\begin{equation}
\label{Solution_m_Four_Dimensional}
m = \frac{R_+^2 +1 - \left(1 \pm i \, R_+ \right) \coth \left( \delta_1 + \delta_2 \right)}{R_+ \left(c_1^2 + s_1^2 - c_2^2  - s_2^2 \right)\mp 2 \, i \, s_1 \, c_1} \, ,
\end{equation}
where $R_+$ is treated as a parameter, on the same footing as $\delta_1, \delta_2$.

We now plug the expression \eqref{Solution_m_Four_Dimensional} for $m$, together with the supersymmetry conditions  \eqref{susy_cond_4d}, into the different quantities summarized in Subsection~\ref{sec:nonsusysol_4d}.
After these manipulations, we find that the chemical potentials in~\eqref{Properties_Solution_Four_Dim} satisfy the relation
\begin{equation}
\label{Constraint_Chemical_Four_Dimensions}
\beta (1 + \Omega - \Phi_1 - \Phi_3) = \mp \, 2 \, \pi \, i  \, .
\end{equation}
This is completely analogous to the one found in five dimensions. We thus argue that it has the same interpretation as an anti-periodicity condition for the Killing spinor when this is translated around the Euclidean time circle.

Again we can introduce the redefined chemical potentials:
\begin{equation}
\label{Little_Potentials_Four_Dimensions}
\omega = \beta \left(\Omega - \Omega^\star \right) , \qquad \varphi_I = \beta \left(\Phi_I - \Phi_I^\star \right) \,,
\end{equation}
where the BPS values $\Omega^\star,\Phi_I^\star$ were given in \eqref{Chemical_Potential_BPS_Four_Dimensions}.
In terms of these variables,  \eqref{Constraint_Chemical_Four_Dimensions} reads:
\begin{equation}
\label{Constraint_Little_Four_Dimensions}
\omega - \varphi_1 - \varphi_3  = \mp \, 2\, \pi \, i \, .
\end{equation}
We find that the explicit expressions of $\omega,\varphi_I$ are:
\begin{align}\label{chem_pot_small_4d}
\omega & = \frac{4  \pi}{ \Upsilon} \left[c_1 \left(c_2 - 2 \, s_2 \right) + s_1 (s_2 - 2 \,c_2 ) \right] \left[R_+ (c_1^2 - c_2^2 + s_1^2 - s_2^2 ) \mp 2 \, i \, c_1 \, s_1 \right] \, , \nn\\
\varphi_1 & = \varphi_2 =  \frac{4  \pi}{ \Upsilon}  \left(-c_1^2 + 2 \, c_1 \, s_1 + c_2^2 - s_1^2 + s_2^2\right) \left[R_+ (c_1 \, s_2 + c_2 \, s_1 ) \mp i \, \rme^{-\delta_1 - \delta_2} \right]  \, , \nn\\
\varphi_3 & = \varphi_4 =  \frac{4  \pi}{ \Upsilon} \!\left[R_+ (c_1^2 - c_2^2 + s_1^2 - s_2^2 )\mp 2  i \, c_1  s_1 \right] \left[(c_1  c_2 + s_1 s_2 ) - (1 \mp i \, R_+ ) (c_1  s_2 + c_2  s_1 ) \right]  , 
\end{align}
where we introduced:
\begin{align}
\Upsilon &=  2 \, R_+ \left(c_1 \, s_2 + c_2 \, s_1 \right) \left[R_+ \left(c_1^2 - c_2^2 + s_1^2 - s_2^2 \right) \mp 4 \, i \, c_1 \, s_1 \right] - c_1 \, s_2 + c_2 \, s_1 \notag \\
& \quad -2 \, \sinh \left(3 \delta_1 + \delta_2 \right) + \sinh \left(\delta_1 + 3 \delta_2 \right)+\cosh \left(3 \delta_1 + \delta_2 \right)-\cosh \left(\delta_1 + 3 \delta_2 \right) \, .
\end{align}

The conserved quantities in~\eqref{Energy, Angular Momentum, Charges_Four_Dim} satisfy
\begin{equation}
\label{Charge_SUSY_Relation_Four_Dimensions}
E - \Omega^\star  J - 2 \, \Phi^\star_1 \, Q_1 -2\, \Phi^\star_3 \, Q_3 = 0  \, ,
\end{equation}
which as already remarked is purely a consequence of supersymmetry. At this stage the expressions for the charges, as well as the one for the entropy, are complex.

After using \eqref{susy_cond_4d}, \eqref{Change_r_to_R_Four_Dimensions}, the on-shell action~\eqref{On_Shell_Action_Euclidean_Four_Dim} can be written in terms of the chemical potentials $\omega,\varphi_1,\varphi_3$ as:
\begin{equation}
\label{On_Shell_Action_With_Potentials_Four_Dimensions}
I = \frac{1}{2 \, i} \frac{\varphi_1 \, \varphi_3 }{\omega} \, . 
\end{equation}
Notice that this is independent of $\beta$, as also found in the five-dimensional analysis.
Using \eqref{Charge_SUSY_Relation_Four_Dimensions}, the quantum statistical relation~\eqref{Quantum_Statistical_Relation_Four_Dimensional} takes the supersymmetric form:
\begin{equation}
\label{QSR_SUSY_Four_Dimensional}
I = - S -  \omega\, J - 2 \, \varphi_1 \, Q_1 -2\, \varphi_3 \, Q_3   	\, .
\end{equation}

We can now take the BPS limit by sending  $R_+ \to R_	\star$, where $R_\star$ is the map of the BPS horizon position $r_\star$ in \eqref{rstar_4d} under the change of coordinate~\eqref{Change_r_to_R_Four_Dimensions}, that is:
\begin{equation}
R_\star = 2 \, s_1 \, c_1 \, m_\star \tanh \left(\delta_1 + \delta_2 \right) \, .
\end{equation}
 In this limit, the complex expression for $m$ in~\eqref{Solution_m_Four_Dimensional} becomes real and gives $m \to m_\star$.  
The original chemical potentials take the values \eqref{Chemical_Potential_BPS_Four_Dimensions}. 
 We define the BPS values of the redefined chemical potentials $\omega$, $\varphi_1$, $\varphi_3$ as
\begin{equation}
\label{BPS_Little_Potentials_Def_Four_Dimensional}
\omega^\star = \lim_{R_+ \to R_\star} \omega \, , \qquad \varphi_I^\star = \lim_{R_+ \to R_\star} \varphi_I \, .
\end{equation}
By evaluating these quantities from \eqref{chem_pot_small_4d}, we obtain  the finite values:
\begin{align}\label{BPS_chempot_4d}
\omega^\star =\, &  -\frac{16 \, \pi}{ \Theta} \left(\rme^{2 (\delta_1+\delta_2)}-3\right) \Big[4 \, \left(s_1 \, c_1 + s_2 \, c_2 \right) \sqrt{s_1 \, s_2 \, c_1 \, c_2 \left(s_1 \, c_2 + s_2 \, c_1 \right)  \rme^{\delta_1 +\delta_2}  } \notag \\[1mm]
& \pm 4 \, i \, s_1 \, s_2 \, c_1 \, c_2 \, \left(c_1 \, c_2 + s_1 \, s_2 \right) \rme^{\delta_1+\delta_2}   \Big] \, ,\notag \\[2mm]
\varphi_1^{\star} = \, & - \frac{16 \, \pi}{ \Theta} \Big\{\sqrt{s_1 \, s_2 \, c_1 \, c_2  \left(s_1 \, c_2 + s_2 \, c_1 \right)  \rme^{(\delta_1 + \delta_2)}  } \left[(\rme^{4 \delta_2}-3) \,\rme^{2 \delta_1}  - 4  s_2\,  c_2 +2\,  \rme^{-2 \delta_1} \right] \notag \\[1mm]
&   \mp 2 \, i \, s_2 \, c_2 \left[\rme^{2 (\delta_1+\delta_2)} \left(c_1^2 + s_1^2 + 2 s_1 \, c_1 + 2s_2 \, c_2  \right) -c_2^2 - s_2^2 - 4 s_2 \, c_2 \right]\Big\} \, ,
\end{align}
where $\varphi_3^{\star}$ is obtained from $\varphi_1^{\star}$ by switching $\delta_1$ and $\delta_2$, and
\begin{align}
\Theta= \, & \rme^{2 (\delta_1+\delta_2)} \left(\rme^{4 \delta_1}+\rme^{4 \delta_2}-10\right)+6 \, \rme^{4 (\delta_1+\delta_2)}+\rme^{6 (\delta_1+\delta_2)} -2 \left(\rme^{-4 \delta_1}+\rme^{-4 \delta_2} \right) \notag \\
& -2 \, \left[4 \left(\rme^{4 \delta_1}+\rme^{4 \delta_2 }\right)-7\right]+\rme^{-2 (\delta_1+\delta_2)} \left[5 \left(\rme^{4 \delta_1}+\rme^{4 \delta_2}\right)-3\right] \, .
\end{align}
These formulae show that the chemical potentials $\omega,\varphi_1,\varphi_3$ remain  complex even after the BPS limit is taken. 
Since the limit is smooth, these still satisfy the constraint~\eqref{Constraint_Little_Four_Dimensions},
\begin{equation}
\label{Constraint_Little_BPS_Four_Dimensions}
\omega^\star - \varphi^\star_1 -  \varphi^\star_3 = \mp \, 2 \, i \, \pi \, ,
\end{equation}
 the on-shell action at the BPS point reads
\begin{equation}
\label{On_Shell_Action_BPS_4d}
I ^\star= \frac{1}{2 \, i} \frac{\varphi^\star_1 \, \varphi^\star_3}{\omega^\star} \, ,
\end{equation}
and satisfies
\begin{equation}
\label{QSR_SUSY_Four_Dimensional}
I^\star = - S^\star -  \omega^\star J^\star - 2 \, \varphi_1^\star \, Q_1^\star -2\, \varphi_3^\star \, Q_3^\star   	\, .
\end{equation}

The supersymmetric on-shell action \eqref{On_Shell_Action_With_Potentials_Four_Dimensions} matches the entropy function proposed in  \cite{Choi:2018fdc}. It was shown there that the BPS entropy follows from Legendre transforming this entropy function and demanding reality of the Legendre transform.
Here we have provided a derivation of the entropy function from imposing supersymmetry in the black hole thermodynamics. The relation with the entropy is clear from \eqref{On_Shell_Action_BPS_4d}.
The expressions \eqref{BPS_chempot_4d} for the BPS chemical potentials match the saddle point values obtained from Legendre transforming the entropy function, as it can be checked by plugging the formulae \eqref{BPS_charges_4d} for the BPS charges in the saddles given in  \cite{Choi:2018fdc}, and comparing with \eqref{BPS_chempot_4d}.

\section{Rotating, electrically charged AdS$_6$ black holes}\label{6d_section}

\subsection{Properties of the finite-temperature solution}

In this section, we turn to six dimensions, considering the asymptotically AdS$_6$ black hole of~\cite{Chow:2008ip}. This is a solution to the six-dimensional $\CN = \left(1, 0 \right)$, ${\rm SU}(2)$ gauged supergravity of~\cite{Romans:1985tw}, which uplifts to massive type IIA supergravity on $S^4/\mathbb{Z}_2$ \cite{Cvetic:1999un}. The black hole has four conserved quantities: the energy $E$, two angular momenta $J_a, J_b$ and one ${\rm U}(1) \subset {\rm SU}(2)$ electric charge $Q$. Because of the two independent angular momenta, the solution looks slightly different in form from the other solutions considered in this paper. However the BPS limit works in the same way as in the other cases, and is in fact very similar to the original one discussed in \cite{Cabo-Bizet:2018ehj}.

The solution is specified by four parameters $m, \, a, \, b, \, \delta$ and has an outer horizon at $r = r_+$ defined as the largest root of the blackening function 
\begin{equation}
\CR(r) = g^2 \left[ r \left( a^2+r^2 \right) + 2 \, m \, s^2 \right] \left[r \left(b^2+r^2\right)+2\,m\,s^2\right] + \left(a^2+r^2\right) \left(b^2+r^2\right) - 2 \, m \, r \, ,
\end{equation}
where as usual we denote $s = \sinh{\delta}, \, c = \cosh{\delta}$.
The entropy and the chemical potentials of the solution are given by
\begin{align}
\label{Chemical_Potentials_Six}
S & = \frac{2 \, \pi^2 \left[ \left(r_+^2 + a^2 \right) \left(r_+^2 + b^2 \right) + 2 \, m \, r_+\, s^2 \right]}{3 \, \Xi_a \, \Xi_b} \, , \notag \\[1mm]
\Omega_a & = a \, \frac{ \left(1 + g^2 \, r_+^2 \right) \left(r_+^2 + b^2 \right) + 2 \, m \, g^2 \, r_+ \, s^2}{\left(r_+^2 + a^2 \right) \left(r_+^2 + b^2 \right) + 2 \, m \, r_+ \, s^2} \, , \notag \\[1mm]
\Omega_b & = b \, \frac{ \left(1 + g^2 \, r_+^2 \right) \left(r_+^2 + a^2 \right) + 2 \, m \, g^2 \, r_+ \, s^2}{\left(r_+^2 + a^2 \right) \left(r_+^2 + b^2 \right) + 2 \, m \, r_+ \, s^2}  \, , \notag \\[1mm]
\Phi & = \frac{2 \, m \, r_+ \, s \, c}{\left(r_+^2 + a^2 \right) \left(r_+^2 + b^2 \right) + 2 \, m \, r_+ \, s^2}  \,, \notag \\[1mm]
\frac{1}{\beta} & = \frac{2 r_+^2 (1 + g^2 r_+^2 ) (2 r_+^2 + a^2 + b^2 ) - (1 - g^2 r_+^2 ) (r_+^2 + a^2 ) (r_+^2 + b^2 ) + 8 m g^2 r_+^3 s^2 - 4 m^2 g^2 s^4}{4 \, \pi \, r_+ \left[\left(r_+^2 + a^2 \right) \left(r_+^2 + b^2 \right) + 2 \, m \, r_+ \, s^2 \right]   } \, ,
\end{align}
where $\Xi_a = 1 - a^2 \, g^2$ and $\Xi_b = 1 - b^2 \, g^2$. 
The energy, the angular momenta and the electric charge read
\begin{align}
E & = \frac{2 \, \pi \, m}{3 \, \Xi_a \, \Xi_b} \left[\frac{1}{\Xi_a} + \frac{1}{\Xi_b} + s^2 \left(1 + \frac{\Xi_a}{\Xi_b} + \frac{\Xi_b}{\Xi_a} \right) \right],  \, \qquad Q = \frac{2 \, \pi \, m \, s \, c}{\Xi_a \, \Xi_b} \,,\notag \\
J_a & = \frac{2 \, \pi \, m \, a}{3 \, \Xi_a^2 \, \Xi_b} \left(1 + \Xi_b \, s^2 \right) , \, \qquad J_b = \frac{2 \, \pi \, m \, b}{3 \, \Xi_a \, \Xi_b^2} \left(1 + \Xi_a \, s^2 \right) \, .
\end{align}
These satisfy the first law of black hole thermodynamics,
\begin{equation}
\diff E = T \, \diff S + \Omega_a \, \diff J_a + \Omega_b \, \diff J_b + \Phi \, \diff Q \, .
\end{equation}
For the black hole under consideration, the quantum statistical relation reads
\begin{equation}\label{QSR_6d}
I = \beta \, E - S - \beta \, \Omega_a \, J_a - \beta \, \Omega_b \, J_b - \beta \, \Phi \, Q \,.
\end{equation}
While in the four- and five-dimensional cases we explicitly verified the validity of the quantum statistical relation by computing the on-shell action, in the present case we assume its validity and use it to obtain an expression for the on-shell action~$I$. 
We will demonstrate that chemical potentials satisfying the correct complex constraint arise from a suitably complexified family of supersymmetric solutions, and that the expression of $I$ on these solutions is precisely the entropy function given in~\cite{Choi:2018fdc}.

\subsection{The BPS solution}
For ease of computation we set $g=1$ from now on. As discussed in \cite{Chow:2008ip}, the solution is supersymmetric if
\begin{equation}
\label{SUSY_6d}
\rme^{2 \delta} = 1 + \frac{2}{a + b}\,.
\end{equation}
We shall always use this condition to eliminate $\delta$ in all the expressions below. We are thus left with the remaining three free parameters $m$, $a$, $b$. 
The supersymmetric solution is free from closed timelike curves if and only if 
\begin{equation}
m = m_\star = \frac{\left(a + b \right)^2 \left( 1 + a \right) \left(1 + b \right) \left(2 + a + b \right)}{2 \left(1 + a + b \right)} \sqrt{\frac{a \, b}{1 + a + b}} \, .
\end{equation}
Imposing both this condition and~\eqref{SUSY_6d}, the temperature vanishes and we obtain the BPS solution. The BPS horizon is located at
\begin{equation}
\label{r_star_six}
r_\star = \sqrt{\frac{a \, b}{1 + a + b}} \, .
\end{equation} 
In the BPS solution, the chemical potential take the BPS values
\begin{equation}
\Omega^\star_a = \Omega^\star_b = 1 \, , \qquad \Phi^\star = 1 \, , \qquad \beta \to \infty \, ,
\end{equation}
while the BPS charges are: 
\begin{align}
E^\star & = -\frac{\pi \,  r_\star \left(a+b \right) \left[2 \, a^2 + a \, \left(b-1 \right) + \left(b+1 \right) \left(2 b-3 \right) \right]}{3 \left(a-1 \right)^2 \left(b-1 \right)^2 \left(a+b+1 \right) } \, , \notag \\
J_a^\star & = -\frac{\pi \, r_\star^3 \left(a+b \right) \left( a + 2 \, b + 1 \right)}{3 \, b \left( a-1 \right)^2 \left(b-1 \right)}  \, , \notag \\
J_b^\star & = -\frac{\pi \,  r_\star^3  \left( a + b \right) \left(2 \, a + b + 1 \right)}{3 \, a \left(a - 1 \right) \left(b - 1 \right)^2} \, , \notag \\
Q^\star & = \frac{\pi \, r_\star \left( a + b \right) }{ \left( a - 1 \right) \left( b - 1 \right) } \, \notag \, .
\end{align}
The quantity above satisfy the supersymmetry relation
\begin{equation}
E^\star - \Omega_a^\star \, J_a^\star - \Omega_b^\star \, J_b^\star - \Phi^\star \, Q^\star = 0 \, .
\end{equation}
The BPS entropy of the BPS black hole solution reads
\begin{equation}
\label{Entropy_Star_Six}
S^\star= \frac{2 \, \pi^2  \, r_\star^2 \, \left(a + b \right)}{3 \left(1 - a \right) \left(1 - b \right) }  \, ,
\end{equation}
and satisfies the following two relations, which involve also the BPS charges \cite{Choi:2018fdc}
\begin{align}
& S^{\star \, 3} - \frac{2 \, \pi^2}{3} S^{\star \, 2} - 12 \, \pi^2 \left(\frac{Q^\star}{3} \right)^2 \, S^\star + \frac{8 \, \pi^4}{3} \, J^\star_a \, J^\star_b = 0 \, , \notag \\
& \frac{Q^\star}{3} S^{\star \, 2} + \frac{2 \, \pi^2}{9} \left(J_a^\star + J_b^\star \right) S^\star - \frac{ 4 \, \pi^2}{3} \left( \frac{Q^\star}{3} \right)^3 = 0 \, .
\end{align}
These can be used to express the BPS entropy in terms of the charges and to obtain a relation between $J_a^\star, \, J_b^\star \, , Q^\star$, analogously to what happens in the other spacetime dimensions.

In the following we assume $0<a <1$, $0<b<1$, which guarantee $r_\star$ to be real and the BPS entropy to be real and positive.

\subsection{The BPS limit}

As in the previous cases, the complexified family of supersymmetric solutions is obtained by solving the equation $\mathcal{R}(r_+) = 0$ for the parameter $m$, so that this is traded for the position of the outer horizon $r_+$. The equation is already of quadratic order in $m$, therefore we can solve it without changing the radial coordinate. Doing so, we obtain
\begin{equation}
m = \frac{1}{2} \left(r_+ \mp i \right) \left( a \pm i \, r_+ \right) \left(b \pm i \,  r_+ \right) \left( a + b \right) \left( a + b + 2 \right) \, ,
\end{equation}
so that $m$ is complex for real values of $a, b, r_+$.
Plugging this expression for $m$ in the chemical potentials~\eqref{Chemical_Potentials_Six}, we find that these satisfy the constraint
\begin{equation}
\beta \left(1 + \Omega_a + \Omega_b - 3 \,  \Phi \right) = \mp 2 \, \pi\, i \, .
\end{equation}
It follows that the redefined chemical potentials
\begin{equation}\label{redef_pot_6d}
\omega_a = \beta \left(\Omega_a - \Omega^\star_a \right) \, , \qquad \omega_b = \beta \left(\Omega_b - \Omega^\star_b \right) \, , \qquad \varphi = \beta \left(\Phi - \Phi^\star \right) \, ,
\end{equation}
 satisfy
\begin{equation}
\omega_a + \omega_b - 3 \, \varphi = \mp \, 2 \, \pi \, i \, .
\end{equation} 
The black hole charges satisfy the supersymmetry condition
\be\label{susycharges_6d}
E - \Omega_a^\star \, J_a - \Omega_b^\star \, J_b - \Phi^\star \, Q = 0 \, .
\ee
Using \eqref{redef_pot_6d}, \eqref{susycharges_6d} in \eqref{QSR_6d}, we obtain the supersymmetric quantum statistical relation
\begin{equation}
I = - S - \omega_a \, J_a - \omega_b \, J_b - \varphi \, Q \, .
\end{equation}
Evaluating $I$ from this expression, we find
\begin{equation}
I = \frac{\pi\,i}{3} \, \frac{\varphi^3}{\omega_a \, \omega_b} \, ,
\end{equation}
which reproduces the entropy function proposed in  \cite{Choi:2018fdc}.

We now take the extremal limit by sending $r_+\to r_\star$. The chemical potentials take the limiting values 
\begin{align}
\omega_a^\star & = \frac{2 \, i \, \pi  \left( a - 1 \right) \left(  b + i \, r_\star \right)}{2 \, i \, a \, b \, r_\star^{-1} + a \, b + a + b - 3 \, r_\star^2} \, , \notag \\ 
\omega_b^\star & =   \frac{2 \, i \, \pi  \left( b - 1 \right) \left(  a + i \, r_\star \right)}{2 \, i \, a \, b \, r_\star^{-1} + a \, b + a + b - 3 \, r_\star^2} \, , \notag \\ 
\varphi & = \frac{2 \, i \, \pi \left(a + i \, r_\star \right) \left(b+i \, r_\star \right)}{2  \, i \, a \, b \, r_\star^{-1} + a \, b + a + b -3 r_\star^2} \, . 
\end{align}
Since the limit is smooth, these still satisfy the constraint
\begin{equation}
\omega^\star_a + \omega^\star_b - 3 \, \varphi^\star = \mp 2 \, \pi \, i \, , 
\end{equation}
and the BPS on-shell action is 
\begin{equation}
I^\star = \frac{\pi\, i}{3} \, \frac{(\varphi^\star)^3}{\omega^\star_a \, \omega^\star_b} \, . 
\end{equation}
We obtained in this way a derivation of the BPS AdS$_6$ black hole entropy function.

\section{Rotating, electrically charged AdS$_7$ black holes}\label{7d_section}

Finally, we discuss asymptotically AdS$_7$ black holes. We consider the seven-dimensional solution originally found in \cite{Chong:2004dy} and further discussed in \cite{Cvetic:2005zi}, and we study its BPS limit. Since this goes through in the same way as in the previous cases, we will keep the presentation short.

\subsection{Properties of the finite-temperature solution}

The seven-dimensional black hole discussed in \cite{Cvetic:2005zi} is a solution to maximal $\SO(5)$ gauged supergravity, and uplifts to eleven-dimensional supergravity on $S^4$~\cite{Cvetic:1999xp}.
We start with a brief summary of the relevant properties of the finite-temperature solution.\footnote{We correct a few misprints in \cite{Cvetic:2005zi} following \cite{Hosseini:2018dob,Choi:2018hmj}.}
This is controlled by four parameters $m$, $a$, $\delta_1$, $\delta_2$ and is given in terms of the following functions:
\begin{align}
& s_I = \sinh{\delta_I} \, , \quad c_I = \cosh{\delta_I} \, , \quad \Xi_{\pm} = 1 \pm a \, g \, , \quad \Xi = 1 - a^2 \, g^2 \, , \quad \rho = \sqrt{\Xi} \, r \, , \notag \\
& H_I = 1 + \frac{ 2 \, m \, s_I^2}{\rho^4}  \, , \quad \alpha_1 = c_1 - \frac{1}{2} \left(1 - \Xi_+^2 \right) \left(c_1 - c_2 \right) \, , \quad \alpha_2 = c_2 + \frac{1}{2} \left(1 - \Xi_+^2 \right) \left(c_1 - c_2 \right)  , \notag \\ 
& \beta_1 = - a \, \alpha_2 \, , \quad \beta_2 = - a \, \alpha_1 \, , \  
\end{align}
\begin{align}
& f_1 (r) = \Xi  \rho^6 H_1 H_2 - \frac{4 \Xi_+^2 m^2 a^2 s_1^2 s_2^2}{\rho^4} + \frac{m  a^2}{2} \Big[4 \Xi_+^2 + 2  c_1  c_2 (1 - \Xi_+^4 ) + (1 - \Xi_+^2 )^2 (c_1^2 + c_2^2 ) \Big]  , \notag \\
& f_2 (r) = - \frac{1}{2} \, g \, \Xi_+ \, \rho^6 H_1 H_2 + \frac{1}{4} m a \left[2 \left(1 + \Xi_+^4 \right) c_1 c_2 + \left(1 - \Xi_+^4 \right) \left(c_1^2 + c_2^2 \right) \right] \, , \notag \\
& Y(r) = g^2 \rho^8 H_1 H_2 + \Xi \rho^6 + \frac{1}{2} m a^2 \left[4 \Xi_+^2 + 2 \left(1 - \Xi_+^4 \right) c_1 c_2 + \left(1 - \Xi_+^2 \right)^2 \left(c_1^2 + c_2^2 \right) \right] \notag \\
& \qquad- \frac{1}{2} m \rho^2  \Big[ 4 \Xi + 2 a^2 g^2 \left(6 + 8 a g + 3 a^2 g^2 \right) c_1 c_2 - a^2 g^2 \left(2 + a g \right) \left(2 + 3 a g \right) \left(c_1^2 +c_2^2 \right) \Big]  ,
\end{align}
where $r$ is the radial coordinate. The outer horizon is found at $r = r_+$, defined as the largest root of the equation $Y(r)=0$.

The entropy, inverse temperature, angular velocity and electrostatic potentials on the
horizon, measured in a non-rotating frame at infinity, read:
\begin{align}
& S = \frac{\pi ^3 \,  \rho ^2 \, \sqrt{f_1} }{4 \, \Xi^3 } \, , \qquad \beta = T^{-1} = 4 \, \pi \,  g \, \rho^3 \sqrt{\Xi \, f_1} \, \left(\frac{\diff Y}{\diff r}\right)^{-1} \, , \notag \\[1mm]
& \Omega = - \frac{1}{g} \left(g + \frac{2 \, f_2}{f_1} \, \Xi_- \right) \, , \qquad \Phi_I = \frac{4 m s_I}{\rho^4 \Xi H_I} \left(\alpha_I \Xi_- + \beta_I \frac{2 f_2 \Xi_-}{f_1} \right) \, ,
\end{align}
while the energy, angular momentum and electric charges are:
\begin{align}
E & = \frac{m \, \pi^2}{32 \, g \, \Xi^4  }  \Big[ 12 \Xi_+^2 \left(\Xi_+^2 - 2 \right) - 2 c_1 c_2 a^2 g^2 \left( 21 \Xi^4_+ - 20 \Xi^3_+ - 15 \Xi^2_+ - 10 \Xi_+ - 6 \right) \notag \\ 
& \qquad \qquad \qquad + \left(c_1^2 + c_2^2 \right) \left(21 \Xi_+^6 - 62 \Xi_+^5 + 40 \Xi_+^4 + 13 \Xi_+^2 - 2 \Xi_+  + 6 \right) \Big] \, , \notag \\
J & = - \frac{m  a  \pi^2}{16 \, \Xi^4 } \Big[ 4 a g \Xi_+^2 - 2 c_1 c_2 (2 \Xi_+^5 - 3 \Xi^4_+ - 1 ) + a g (c_1^2 + c_2^2 ) (\Xi_+ + 1 ) (2 \Xi^3_+ - 3 \Xi^2_+ - 1 ) \Big] \, , \notag \\ 
Q_1 \!&  = \frac{m \, s_1 \, \pi^2}{8 \, g \, \Xi^3 } \Big[a^2 g^2 c_2 \left(2 \Xi_+ + 1 \right) - c_1 \left(2 \Xi_+^3 - 3 \Xi_+^2 - 1 \right) \Big] \, , \notag \\ 
Q_2 \!& = \frac{m \, s_2 \, \pi^2}{8 \, g \, \Xi^3 } \Big[a^2 g^2 c_1 \left(2 \Xi_+ + 1 \right) - c_2 \left(2 \Xi_+^3 - 3 \Xi_+^2 - 1 \right) \Big] \, .
\end{align}
The energy is given in a scheme such that the energy of the vacuum AdS$_7$ solution is $E_0=0$. In a different scheme, the expression above should be regarded as $E-E_0$ (considerations similar to the ones discussed in the five-dimensional case apply here). 
The quantities above satisfy the first law of black hole thermodynamics,
\begin{equation}
\diff E = T \, \diff S + 3 \, \Omega \, \diff J + \Phi_1 \, \diff Q_1 + \Phi_2 \, \diff Q_2 \, .
\end{equation}
The quantum statistical relation reads:
\be\label{QSR_7d}
I = \beta E  -  S - 3\beta\, \Omega \, J - \beta \,  \Phi_1 \, Q_1 - \beta  \, \Phi_2 \,  Q_2  \, .
\ee
As for the six-dimensional case, we will assume this is satisfied without evaluating the on-shell action $I$ independently.
 For consistency with the assumption made for the vacuum energy, we assume we are working in a scheme where $I_0=\beta E_0 =0$.
We now show that the expression of $I$ on the complexified family of solutions coincides with the entropy function of \cite{Hosseini:2018dob}, and that the constraint on the chemical potentials arises in the same way as in the previous sections.

\subsection{The BPS solution}

We will set $g=1$ from now on. The solution is supersymmetric (preserving two supercharges) if \cite{Cvetic:2005zi}
\begin{equation}\label{susycond_7d}
a = \frac{2}{3(1- \rme^{\delta_1+\delta_2})} \, .
\end{equation}
We will always use this relation to eliminate $a$ in the expressions below. The remaining parameters are $m,\delta_1,\delta_2$. For simplicity we will set $\delta_1 =\delta_2\equiv \delta$ (and similarly $c_1=c_2 \equiv c$, $ s_1=s_2\equiv s$), the extension to the case $\delta_1\neq \delta_2$ being straightforward although more involved.
We thus have $\Phi_1 =\Phi_2\equiv \Phi$ and $Q_1 =Q_2 \equiv Q$.

Closed timelike curves are avoided by taking
\begin{equation}
m = m_\star = \frac{4 \, \rme^{-3 \delta} \left(c + 2 \, s \right)^3}{729 \, c^2 \, s^6} \, ,
\end{equation}
imposing this in addition to \eqref{susycond_7d} implies vanishing of the temperature and thus leads to the BPS solution. The BPS horizon is located at
\begin{equation}
\label{r_star_dim}
r^2_\star =- \frac{16}{3 \left(2 \, \rme^{2 \delta} - 3 \, \rme^{4 \delta} + 5 \right)} \, ,
\end{equation}
note that, since $r_\star^2$ should be positive, the equation above implies 
\begin{equation}
\rme^{2\delta} > \frac{5}{3} \,  .
\end{equation}
This is a physical condition on the parameter $\delta$ and we shall assume it in the following.

The chemical potentials take the BPS values
\begin{equation}
\Omega^\star = 1 \, , \qquad \Phi^\star = 2 \, , \qquad \beta \to \infty\, .
\end{equation}
The BPS charges are: 
\begin{align}
E^\star = & \, \frac{16 \, \pi ^2 \left(-21 \, \rme^{4 \delta} + 18 \, \rme^{6 \delta} + 7 \right)}{3 \left(5 - 3 \, \rme^{2 \delta} \right)^4 \left(\rme^{2 \delta} + 1 \right)^2} \, , \nn\\[1mm]
J^\star = & \, \frac{16 \, \pi ^2 \left[ 9 \, \rme^{2 \delta}  \left(\rme^{2 \delta} + 2 \right) - 23 \right]}{9 \left(5 - 3 \, \rme^{2 \delta} \right)^4 \left(\rme^{2 \delta} + 1 \right)^2} \, , \nn\\[1mm]
Q^\star = & \, -\frac{\pi^2  \, \tanh\delta \, \rme^{-3 \delta}}{\left(c - 4 \, s \right)^3} \, ,
\end{align}
and satisfy the supersymmetry relation
\begin{equation}
E^\star - 3 \, \Omega^\star  J^\star - 2 \, \Phi^\star \, Q^\star = 0 \, .
\end{equation}
The BPS entropy reads
\be
S^\star =  \, \frac{2 \, \pi^3 \sqrt{c + 8 \, s}}{3 \, \rme^{4 \delta} \, \sqrt{3 \, c^3} \left(4 \, s - c \right)^3}  \, ,
\ee
and can be written in terms of the charges as \cite{Choi:2018hmj}:
\begin{equation}
 S^\star  = 2  \pi \,\sqrt{ \frac{32 \, (Q^\star)^3 -  3 \, \pi^2 \, (J^\star)^2}{32 \, Q^\star- \pi^2 } }\,.
\end{equation}

\subsection{The BPS limit}

In order to study the complexified family of supersymmetric solutions, we solve the equation $Y(r_+) = 0$ for the parameter $m$, thus trading it for the position of the outer horizon $r_+$. 
This equation is of quadratic order in $m$, therefore it is immediately solved as:
\begin{align}
\label{m_7_dim}
m = & -\frac{c + 2 \, s}{648 \, s^6} \bigg\{\left[3 \, r_+^2 \left(c^2 - 8 \, c \, s + s^2 + 1 \right) + 8 \, \rme^{-2 \delta} \right] \CR \notag \\ 
& \quad + r_+^2 \left(4 \, s - c \right) \left[2 \, r_+^2 \left(7 \left(c^2 + s^2 \right) + 4 \, c \, s - 9 \right) - 2 \rme^{-2 \, \delta} - 18 \right] + 16 \, \rme^{-3 \delta} \bigg\} \, ,
\end{align}
where
\begin{equation}
\CR = \sqrt{4 \, \rme^{-2 \delta} - 2 \, r_+^2 \left[7 \, \left(c^2 + s^2 \right) + 4 \, c \, s - 9 \right] } \, .
\end{equation}
We now show that this square root is imaginary.
Using the expression for $r_\star$ given in~\eqref{r_star_dim}, we can write
\begin{equation}
\CR = \sqrt{4 - \frac{16 \, r_+^2 \left(r_\star^2 - r_\star \, \sqrt{r_\star^2 + 1} + 1 \right)}{r_\star^2}} \, ,
\end{equation}
and using the physical condition $r_+ > r_\star$ it is easy to see that the argument satisfies the inequality
\begin{equation}
4 - \frac{16 \, r_+^2 \left(r_\star^2 - r_\star \, \sqrt{r_\star^2 + 1} + 1 \right)}{r_\star^2} < 4 - \frac{8 \, r_+^2}{r_\star^2} < 0\,,
\end{equation}
 showing that the square root is always imaginary. The expression for $m$ given in~\eqref{m_7_dim} is therefore  complex. This identifies our complexified family of solutions.
 
Plugging the above expression for $m$ in the chemical potentials, we find that these satisfy the constraint
\begin{equation}
\beta \left(1 + 3 \, \Omega - 2 \, \Phi \right) = \mp 2 \, \pi \, i \, .
\end{equation}
Again we can introduce the redefined chemical potentials
\be
\omega = \beta\,(\Omega - \Omega^\star)\, , \qquad \varphi = \beta\,(\Phi - \Phi^\star)\,.
\ee
We checked that the on-shell action is given in terms of these variables by
\begin{equation}
I = - \frac{\pi^3}{128} \, \frac{\varphi^4}{\omega^3} \, .
\end{equation}
The BPS on-shell action and the BPS chemical potentials satisfy the supersymmetric quantum statistical relation
\begin{equation}
I = - S - 3 \, \omega \, J - 2 \, \varphi \, Q \, .
\end{equation}

At this point we take the extremal limit by sending $r_+\to r_\star$.
The limiting values of the chemical potentials are: 
\begin{align}
\omega^\star & = -\frac{6 \, \pi}{\left(c + 8 \, s \right) \, \Theta}\, \sqrt{2 \left(16 \, c \, s + c^2 + s^2 + 1 \right) }  \left(c - 4 \, s \right) \bigg[c \left(\sqrt{3} - \sqrt{-8 \, \tanh\delta - 1} \right) \notag \\ 
& \qquad + 4 \, s \left(\sqrt{-8 \tanh\delta - 1} +2 \, \sqrt{3} \right) \bigg] \, ,\nn \\
\varphi^\star & = - \frac{64 \, \pi} {\Theta \, \rme^{\delta} \sqrt{2 \left(c^2 + 16 \, c  \, s + s^2 + 1 \right)}} \left(\sqrt{3} \, \rme^{- \delta} \left(c + 8 \, s \right) + 9 \, c \, s  \sqrt{-8 \tanh\delta - 1} \right) \, ,
\end{align}
where we have defined
\begin{align}
\Theta & =  8 \, c \, s \left(\sqrt{-3 \left(8 \, \tanh\delta + 1 \right)} - 18 \right) + \left(23 \sqrt{-3 \left(8 \, \tanh\delta + 1 \right)}-9\right) \left(c^2 + s^2 \right) \notag \\ 
& \quad -9 \left(\sqrt{-3 \left(8 \, \tanh\delta + 1 \right)}+1 \right) \, .
\end{align}
These BPS chemical potentials satisfy the constraint
\begin{equation}
3 \, \omega^\star - 2 \, \varphi^\star = \mp \, 2 \, \pi \, i \, ,
\end{equation}
which is completely analogous to what we have found in lower dimensions.
In terms of these chemical potentials, the on-shell action reads
\begin{equation}
I^\star = - \frac{\pi^3}{128} \, \frac{\left(\varphi^\star \right)^4}{\left(\omega^\star \right)^3} \, .
\end{equation}
This completes our derivation of the BPS entropy function from black hole thermodynamics.

\section{Discussion}\label{sec:conclusions}

The main result of this paper has been to extend the BPS limit of rotating AdS black hole thermodynamics defined in \cite{Cabo-Bizet:2018ehj} to five-dimensional solutions with more than one electric charge, as well as to other spacetime dimensions. In each case, we have provided a derivation of the extremization principle leading to the Bekenstein-Hawking entropy, by showing that the entropy functions of \cite{Hosseini:2017mds,Hosseini:2018dob,Choi:2018fdc} are the supergravity action $I = I(\omega,\varphi)$ evaluated on a complexified family of supersymmetric solutions, and that the supersymmetric chemical potentials $\omega,\varphi$ indeed satisfy the corresponding extremization equations.

The analysis of several examples across different spacetime dimensions demonstrates that this approach is general and should play a role towards understanding the thermodynamics of BPS black holes in AdS. 
 As summarized in Section~\ref{sec:introduction}, the gravitational analysis also defines the microscopic, dual SCFT partition function which should reproduce the entropy function in the large $N$ regime. Progress in this direction has been made recently, and the entropy function for rotating BPS black holes in AdS$_5$ has been obtained from a dual SCFT$_4$ by different methods:
from a version of the supersymmetric Casimir energy in \cite{Cabo-Bizet:2018ehj}, from a Cardy-like limit of the superconformal index in \cite{Choi:2018hmj,Honda:2019cio,ArabiArdehali:2019tdm,Kim:2019yrz,Cabo-Bizet:2019osg,Amariti:2019mgp} (see also \cite{Choi:2019miv} for an AdS$_6$/CFT$_5$ study), and from the large $N$ limit of the $\mathcal{N}=4$ SYM index in~\cite{Benini:2018ywd}.
The AdS$_4$/CFT$_3$ case is also interesting: the class of black holes studied in this paper uplifts to eleven-dimensional supergravity on $S^7$, hence the dual field theory is the ABJM theory on $S^1\times S^2$, with an anti-periodic supercharge and chemical potentials satisfying the complex constraint. It should thus be possible to retrieve the entropy function by evaluating the partition function \eqref{SCFT_Z} for the ABJM theory at large $N$.

On the gravity side, there are some generalizations of our work that it would be interesting to consider. Our analysis of the solutions of \cite{Cvetic:2005zi,Chow:2008ip} strongly indicates that the same BPS limit will work when the most general set of electric charges and angular momenta is turned on in each spacetime dimension, although in many cases the corresponding asymptotically AdS black hole solutions are still to be constructed, and the explicit check may be technically hard to perform.
Specifically, in the context of eleven-dimensional supergravity on $S^7$, one could relax the condition of pairwise equal electric charges within the ${\rm U}(1)^4$ consistent truncation of SO(8) maximal supergravity, though the finite-temperature asymptotically AdS$_4$ solution with four independent electric charges in addition to the angular momentum has not been found yet.\footnote{Very recently, the corresponding BPS solution has been constructed in \cite{Hristov:2019mqp}.} In four dimensions, one could also switch on magnetic charges. For type IIB supergravity on $S^5$, one could analyze the solution carrying two independent angular momenta and three independent electric charges given in \cite{Wu:2011gq}. For massive type IIA supergravity on $S^4/\mathbb{Z}_2$, the solution of \cite{Chow:2008ip} discussed here already carries all possible independent electric charges and angular momenta available within known consistent truncations, although one may still search for asymptotically AdS$_6$ black holes carrying a non-vanishing electric charge for the additional ${\rm U}(1)\subset {\rm SU}(2)$ isometry of $S^4$  working directly in ten dimensions (this charge would be dual to a flavor charge of the D4-D8-O8 SCFT$_5$). For eleven-dimensional supergravity on $S^4$, a solution carrying two independent electric charges and three independent angular momenta is likely to exist within the ${\rm U}(1)^2$ truncation of seven-dimensional {\rm SO}(5) maximal supergravity, but has not been found yet. Nevertheless, the known solutions allow to partially relax the condition of equal electric charges and equal angular momenta we imposed: one could take the two electric charges in the solution of \cite{Cvetic:2005zi} to be independent, or consider the solution with equal electric charges but independent angular momenta given in~\cite{Chow:2007ts}.

 We would like to comment further on five-dimensional black holes, going beyond $S^5$ compactifications of type IIB supergravity. The 
  imprint of $S^5$ in the five-dimensional supergravity considered in this paper is found in the specific number of vector multiplets (three, gauging the $\U(1)^3\subset\SO(6)$ isometry group of $S^5$) and in the form of the $C_{IJK}$ tensor controlling the matter couplings (see Appendix \ref{app:OnShActionsAdS5} for details). 
Multi-charge, supersymmetric AdS$_5$ black holes are known more generally in five-dimensional Fayet-Iliopoulos gauged supergravity with an arbitrary number $n_V$ of vector multiplets and an arbitrary choice of the tensor $C_{IJK}$,  $I
,J,K=1,\ldots,n_V+1$ (under the assumption that the scalar manifold is symmetric) \cite{Gutowski:2004yv,Kunduri:2006ek}.  The solutions carry angular momenta $J_1^\star,J_2^\star$ and $n_V+1$ electric charges $Q_I^\star$.
 An entropy function whose Legendre transform should reproduce the Bekenstein-Hawking entropy of these black holes has been conjectured in \cite[Appendix A]{Hosseini:2018dob} and reads
 \be\label{entropyfctSE}
 I = \frac{\pi}{24}\,\frac{C_{IJK}\varphi^I \varphi^J \varphi^K}{\omega_1\,\omega_2}\,. 
 \ee
In Appendix~\ref{app:Leg_transf} we prove that this is indeed true, provided the chemical potentials satisfy the constraint 
    \be\label{constraintFIsugra}
  \omega_1+\omega_2 - 3\bar{X}_I \varphi^{I} = \mp\, 2\pi i\,,
  \ee
   and in addition one demands reality of the Legendre transform.
Here we give the \hbox{saddle} point expressions for the BPS chemical potentials of the black holes in \cite{Gutowski:2004yv}, which carry one angular momentum $ J_1^\star = J_2^\star$ and $n_V+1$ electric charges $Q_I^\star$. 
  Generalizing our formulae \eqref{BPS_Little_Potentials}, we infer that these BPS chemical potentials read: 
\begin{align}\label{conjecture}
\omega_1^\star = \omega_2^\star \equiv \tfrac{1}{2}\,\omega^\star & = - \frac{\pi}{1+\alpha_1 }\bigg[ \frac{ \alpha_2 }{  \sqrt{ 4 \left(1+\alpha_1 \right)\alpha_3  - \alpha_2^2  }} \pm  i \bigg]  , \notag \\[3mm] 
\varphi^{I\,\star} &=    \frac{9\pi}{1+\alpha_1  }C^{IJK}q_J\bigg[  \frac{\alpha_2\,\bar X_K - (1+\alpha_1)\,q_K}{\sqrt{ 4 \left(1+\alpha_1 \right)\alpha_3  - \alpha_2^2  }}     \pm i \,\bar X_K \bigg]  \, ,
\end{align}
where $\bar{X}_K$ are the values taken by the scalar fields in the supersymmetric AdS$_5$ solution, while the real parameters $q_I$, $\alpha_1=\frac{27}{2}C^{IJK}\bar X_I \bar X_J q_K$, $\alpha_2= \frac{27}{2}C^{IJK}\bar X_I q_J q_K$, $\alpha_3=\frac{9}{2}C^{IJK}q_Iq_Jq_K$  control the solution in \cite{Gutowski:2004yv}.
In Appendix~\ref{app:Leg_transf}  we show that the expressions \eqref{conjecture} are indeed saddles of the extremization problem leading to the black hole entropy.   
  
 The entropy function \eqref{entropyfctSE} has been reproduced from a dual SCFT$_4$ viewpoint by taking the Cardy-like limit of the  superconformal index in \cite{Amariti:2019mgp}.
 Some of the black hole solutions of \cite{Gutowski:2004yv,Kunduri:2006ek} may uplift to type IIB supergravity on Sasaki-Einstein manifolds and thus have an SCFT$_4$ dual, however  the uplift is only known for the case of $S^5$, or in a single-charge limit where the black holes are solutions to minimal gauged supergravity. This is in part due to the fact that a consistent truncation of type IIB supergravity on five-dimensional Sasaki-Einstein manifolds including all Kaluza-Klein vector fields gauging the relevant internal symmetries has not been found to date. For these reasons, the details of the matching with the dual SCFT are not fully under control yet. In addition, the
  non-extremal solutions which are the starting point of our limiting procedure are only known within the $\U(1)^3$ theory discussed here. Clearly it would be interesting to construct new asymptotically AdS black holes in five-dimensional supergravity (for instance relaxing the assumption that the scalar manifold is symmetric), study their uplift to type IIB supergravity on different Sasaki-Einstein manifolds, and extend the results of the present paper by investigating their BPS limit.
 Similar considerations can be made for black holes in spacetime dimension different than five. 
  
 Although constructing the full asymptotically AdS$_d$ solution would be desiderable, for the purpose of studying the extremization principle it may be sufficient to focus on the simpler near-horizon geometry, upon identifying the near-horizon counterpart of our BPS limit. This approach, once promoted to the full ten- or eleven-dimensional supergravity theory, may also lead to a generalization of the extremization principle of \cite{Gauntlett:2019roi,Hosseini:2019ddy,Kim:2019umc} to the case of rotating horizons with no magnetic charge.
 
In five dimensions, there are also recently found black holes that are just asymptotically {\it locally} AdS$_5$, since the $S^3$ spatial part of the conformal boundary is squashed \cite{Blazquez-Salcedo:2017ghg,Cassani:2018mlh,Bombini:2019jhp}. It has been shown in \cite{Cassani:2018mlh} that the expression of the Bekenstein-Hawking entropy of these black holes in terms of the charges is the same as in the round $S^3$ case, provided one uses the Page electric charges of the solution. Hence the entropy function should also be the same, provided the electric potentials $\varphi^I$ are those conjugate to the Page charges. It would be interesting to show this from the on-shell action by implementing the BPS limit discussed here.

\section*{Acknowledgments}

We would like to thank Francesco Benini, Nikolay Bobev, Guillaume Bossard and Chiara Toldo for interesting discussions.
DC would like to thank Alejandro Cabo-Bizet, Dario Martelli and Sameer Murthy for the collaboration that led to~\cite{Cabo-Bizet:2018ehj} and comments on the manuscript.

\appendix

\section{The on-shell action of AdS$_5$ black holes}\label{app:OnShActionsAdS5}

In this appendix, we evaluate the holographic charges and the on-shell action of the AdS$_5$ black holes studied in Section~\ref{5d_section}. In order to remove the divergences due to the infinite spacetime volume we adopt holographic renormalization and use the counterterms for Fayet-Iliopoulos gauged supergravity given in \cite{Cassani:2018mlh}.

In this paper we  fix the five-dimensional Newton constant as $G = 1$. The Lorentzian metric has mostly plus $(-,+,\ldots,+)$ signature. In $d$ dimensions, the Hodge star is defined as $\star (\diff x^{\mu_1}\wedge \cdots \wedge \diff x^{\mu_k}) = \frac{1}{(d-k)!}\epsilon^{\mu_1\ldots \mu_k}{}_{\mu_{k+1}\ldots\mu_d}\diff x^{\mu_{k+1}}\wedge \cdots \wedge \diff x^{\mu_d}$, with $\epsilon_{01\ldots(d-1)}=\sqrt{|\det g_{\mu\nu}|}\,$.

The bosonic action of five-dimensional $\mathcal{N}=2$ supergravity coupled to $n_V$ vector multiplets and with Fayet-Iliopoulos gauging is:
\begin{equation}
\label{Bulk_action}
\CS = \frac{1}{16  \pi} \!\int{ \!\Big[ \left(R - 2 \CV \right)\star\! 1  - Q_{IJ}  \diff  X^I\! \wedge \star \diff X^J - Q_{IJ} F^I\! \wedge \star F^J - \frac{1}{6} C_{IJK} F^I\! \wedge F^J \!\wedge A^K \Big] },
\end{equation}
where $A^I$ are Abelian gauge fields with field strength $F^I=\diff A^I$, $I=1,\ldots,n_{V}+1$,  $C_{IJK}$ is a constant symmetric tensor, and the real scalar fields $X^I$ are subject to the constraint
\begin{equation}
\label{constraint}
\frac{1}{6}\, C_{IJK} X^I X^J X^K = 1\,,
\end{equation}
so that there are only $n_V$ dynamical scalars.
The kinetic matrix $Q_{IJ}$ is given by
\begin{equation}
\label{Qmatrix}
Q_{IJ} = \frac{9}{2}X_I  X_J - \frac{1}{2} \, C_{IJK} X^K\,,
\end{equation}
where the lower-index scalar fields $X_I$ are defined as
\begin{equation}
\label{def_X_I}
X_I=\frac{1}{6} C_{IJK} X^J X^K.
\end{equation}

We assume that the scalar fields parameterize a symmetric space. In this case the $C_{IJK}$ tensor satisfies the identity 
\be\label{3Cinto1C}
C_{IJK} C_{J'(LM} C_{PQ)K'} \,\delta^{JJ'}\delta^{KK'}   \,=\, \frac{4}{3}\,\delta_{I(L} C_{MPQ)}\,.
\ee
This condition is satisfied by the $\U(1)^3$ theory we will focus on momentarily. The same assumption was made in \cite{Gutowski:2004yv,Kunduri:2006ek} to construct general supersymmetric black hole solutions to Fayet-Iliopoulos gauged supergravity. 

Denoting by $\bar{X}^I$ the constant values of the scalars $X^I$ in the supersymmetric AdS$_5$ vacuum, 
 the scalar potential can be written as
\begin{equation}
\label{Scalar_Potential}
\CV = -6 \, g^2 \, \bar{X}^I \, X_I\ ,
\end{equation}
where $g$ is a coupling constant keeping track of the gauging (the Fayet-Iliopoulos gauging parameters $V_I$ are related to the $\bar{X}^I$ as $ V_I = g\bar{X}_I$; therefore the $\bar{X}^I$ should be seen as parameters of the Lagrangian and not of the solution).
Given the real superpotential function
\begin{equation}
\label{Superpotential}
\CW = 3 \, g \, \bar{X}_I \, X^I 
\end{equation}
 appearing in the supersymmetry variation of the gravitino, the scalar potential can be obtained via the formula:
\begin{equation}
\CV = \frac{1}{2} \Big(Q^{IJ} - \frac{2}{3} \, X^I \, X^J \Big) \, \frac{\partial \CW}{\partial X^I} \, \frac{\partial \CW}{\partial X^J} - \frac{2}{3} \, \CW^2\, .
\end{equation}

Our spacetime $M$ can be seen as a foliation of co-dimension one hypersurfaces of constant $r$. We  denote the hypersurfaces by $\partial M_r$, while $M_r$ will be the interior region bounded by $\partial M_r$. The metric 
 has the form:
\begin{equation}
\label{Induced_Metric}
\diff s^2_5 = g_{rr} \, \diff r^2 + h_{ij} (r,x ) \,\diff x^i \, \diff x^j \, ,
\end{equation} 
where $i,j = 0, \dots, 3$ and $h_{ij} ( r, x )$ is the induced metric on $\partial M_r$.
In order to regulate the large-distance divergences, we impose a cutoff $\rnot$, so that the solution extends only up to $r=\rnot$.
 Holographic renormalization introduces suitable local counterterms on $\partial M_{\rnot}$ which cancel the divergences appearing in the supergravity action for $\rnot \to \infty$. This yields the renormalized action 
\begin{equation}
\CS_{\text{ren}} = \lim_{\rnot \to \infty}  \, \CS_{\text{reg}} \, ,
\end{equation}
where the regulated action $S_{\text{reg}}$ is the sum of three pieces:
\begin{equation}
\CS_{\text{reg}} = \CS_{\text{bulk}} + \CS_{\text{GH}} + \CS_{\text{ct}} \, .
\end{equation}
Here, $\CS_{\text{bulk}}$ is the bulk supergravity action \eqref{Bulk_action} evaluated on the regulated spacetime $M_{\rnot}$. Using the trace of the Einstein equation as well as the Maxwell equation this can be put in the form:
\begin{equation}
\label{Bulk_Action_Intermediate}
\CS_{\text{bulk}} = - \frac{1}{12 \, \pi} \int_{M_{\rnot}}  \, \CV \star1 - \frac{1}{24 \, \pi} \int_{M_{\rnot}} { \diff \left(A^I \wedge Q_{IJ} \star F^J \right) } \, .
\end{equation}
 $\CS_{\text{GH}}$ is the Gibbons-Hawking boundary integral, 
\begin{equation}
\label{Gibbons-Hawking action}
\CS_{\text{GH}} = \frac{1}{8  \pi} \, \int_{\partial M_{\rnot}} {\diff^4 \, x \, \sqrt{h} \, K} \, ,
\end{equation}
where $h = \abs{\det{h_{ij}}}$,  and $K = h^{ij} \, K_{ij}$ is the trace of the extrinsic curvature tensor $K_{ij} = \frac{1}{2 \sqrt{g_{rr}}}  \, \frac{\partial h_{ij}}{\partial r} \, . $
 Finally, the counterterm action $\CS_\text{ct}$ is given by:
\begin{equation}
\label{Counterterm_action_simp}
\CS_\text{ct} =  -\frac{1}{8 \pi} \int_{\partial M_{\rnot}} \diff^4 x \, \sqrt{h} \left( \CW \, + \Xi \, R \right) \, ,
\end{equation}
where $R$ is the Ricci scalar of the induced metric $h_{ij}$ and
the function $\Xi$ is \cite{Cassani:2018mlh}: 
\begin{equation}
\label{Xi_function}
\Xi = \frac{1}{4 \, g} \, X_I \, \bar{X}^I \, .
\end{equation}
In $\CS_\text{ct}$ we have omitted terms involving $\log \rnot$ as they vanish on asymptotically AdS solutions, such as those of interest in this paper.

The holographic energy-momentum tensor is defined as:
\begin{align}
\label{Stress_Energy_Tensor_Formula}
\langle T_{ij} \rangle &= - \lim_{\rnot \to \infty} \frac{2 \rnot^2 g^2}{\sqrt{h}} \, \frac{\delta \CS_\text{reg}} {\delta h^{ij}} \nn\\[1mm]
&= \frac{1}{8 \pi} \lim_{\rnot \to \infty} \rnot^2 g^2 \Big[ -K_{ij} + K\, h_{ij} - \CW\,  h_{ij} + 2\,\Xi \Big(R_{ij} - \frac{1}{2}  R \, h_{ij} \Big)  \Big] \, ,
\end{align}
where $R_{ij}$ and $R$ are the Ricci tensor and the Ricci scalar of $h_{ij}$, respectively. 
The holographic currents sourced by the boundary values of the gauge fields $A^I_i$ are:
\begin{align}\label{Electric_Currents_Formula}
\langle j^i_I \rangle &= \lim_{\rnot \to \infty} \frac{\rnot^4 g^4} {\sqrt{h}} \, \frac{\delta \CS_\text{reg}}{\delta A^I_i} \nn\\[1mm]
&= - \frac{1}{48 \, \pi} \lim_{\rnot \to \infty} \rnot^4 g^4 \left[\epsilon^{ijkl} \left(Q_{IJ} \star F^J + \tfrac{1}{6} C_{IJK} A^J \wedge F^K \right)_{jkl} \right] \, .
\end{align}
Again in $\langle T_{ij} \rangle$ and  in $\langle j^i_I \rangle$ we are omitting the contribution arising from logarithmic terms in $\CS_{\rm ct}$, as their variation vanishes in the solution of interest. We can also introduce the one-point function of the scalar operators sitting in the same supermultiplet as the holographic currents:
\begin{align}
\label{One_Point_Function_Def}
\langle \CO_I \rangle &= \lim_{\rnot \to \infty} \left( \rnot^2  g^2 \log{\left(\rnot^2  g^2 \right)} \, \frac{1}{\sqrt{h}} \, \frac{\delta \CS_\text{reg}}{\delta X^I}  \right) \nn\\
&= \frac{1}{4 \, \pi} \, \bar{Q}_{IJ} \,   \varphi^{J\,(0)}\,,
\end{align}
where $\bar{Q}_{IJ}$ is the kinetic matrix evaluated on the supersymmetric AdS$_5$ vacuum, and 
$\varphi^{I\,(0)}$ is the coefficient of the $\mathcal{O}(r^{-2})$ term in the expansion of the scalar fields $X^I$. We refer to \cite{Cassani:2018mlh} for details on the derivation of these quantities in general asymptotically locally AdS$_5$ solutions to Fayet-Iliopoulos gauged supergravity.

\bigskip

We want to apply the formulae above to the $\U(1)^3$ theory discussed in Section~\ref{5d_section}. This is obtained by setting $n_{V} = 2$ and $C_{IJK}  = |\epsilon_{IJK}| $, where $\epsilon_{IJK}$ is the totally anti-symmetric symbol.
Moreover the Fayet-Iliopoulos gauging parameters are fixed by stating that in the supersymmetric AdS$_5$ vacuum the scalars take the equal values: 
\begin{equation}
\bar{X}^I = 1 \qquad \Rightarrow \qquad \bar{X}_I = \frac{1}{3}\ .
\end{equation}
With these choices, eqs.~\eqref{constraint}--\eqref{Scalar_Potential} specialize to:
\begin{align}
 X^1& \, X^2 \, X^3 = 1\,,\nn\\[1mm]
X_I &= \frac{1}{3} \left(X^I \right)^{-1} \,,\nn\\[1mm]
Q_{IJ} & 
 = \frac{1}{2} \, \text{diag} \, \Big( \left(X^1 \right)^{-2} , \, \left(X^2 \right)^{-2} , \, \left(X^3 \right)^{-2} \Big)\,,\nn\\[1mm]
\CV &= - 2 \, g^2 \, \sum_{I=1}^3\left(X^I \right)^{-1} \,. \label{Scalar_Potential_U1}
\end{align}
Plugging these expressions in \eqref{Bulk_action}, we retrieve the action of the $\U(1)^3$ model given in  \eqref{CGLP_Lagrangian}.
The superpotential \eqref{Superpotential} and the function \eqref{Xi_function} entering in the holographic counterterms read:
\begin{align}
\CW &= g \left(X^1+ X^2 + X^3 \right) \, .\nn\\[1mm]
\Xi & = \frac{1}{12 \, g} \, \left[ \left(X^1 \right)^{-1} + \left(X^2 \right)^{-1} + \left(X^3 \right)^{-1} \right] \, .\label{Superpotential_Xi}
\end{align}

We are now in the position of computing the holographic quantities for the solution reviewed in Section~\ref{sec:nonsusy_5d_sol}.
Due to the symmetries of the solution, the holographic energy-momentum tensor can be written as
\begin{equation}
\label{Stress_Energy_Tensor_Form}
\langle T_{ij} \rangle \, \diff x^i \, \diff x^j = \langle T_{tt} \rangle \, \diff t^2 + \langle T_{\theta \theta} \rangle \, \left(\sigma_1^2 + \sigma_2^2 \right) + \langle T_{\psi \psi} \rangle \, \sigma_3^2 + 2 \, \langle T_{t \psi} \rangle \, \diff t \, \sigma_3 \, ,
\end{equation}
and \eqref{Stress_Energy_Tensor_Formula} gives for its components:
\begin{align}
\label{Stress_Energy_Tensor_Components}
\langle T_{tt} \rangle & = \frac{8 \, g^3 \, m \left(a^2  g^2 + 2 \, s_1^2 + 2 \, s_2^2+2 \, s_3^2 + 3 \right)+3g}{64 \, \pi } \, , \notag \\[1mm]
\langle T_{t \psi} \rangle & = \frac{a \, g^3 \, m \left(s_1 \, s_2 \, s_3 -  c_1 \, c_2 \, c_3 \right)}{4 \, \pi } \, , \notag \\[1mm]
\langle T_{\theta \theta} \rangle & = \frac{8 \, g^2 \, m \left(-3 \, a^2  g^2 + 2 \, s_1^2 + 2 \, s_2^2 + 2 \, s_3^2 + 3 \right)+3}{768 \, \pi \,  g} \, ,\notag \\[1mm]
\langle T_{\psi \psi} \rangle & = \frac{8 \, g^2 \, m \left(9 \, a^2 g^2 + 2 \, s_1^2 + 2 \, s_2^2 + 2 \, s_3^2 + 3 \right)+3}{768 \, \pi \,  g} \, .
\end{align}
Evaluating \eqref{Electric_Currents_Formula} we find that the only non-vanishing components of the electric currents are
\begin{align}
\label{Electric_Currents_Components}
\langle j_I^t \rangle & = -\frac{m\, g^3 \, c_I \, s_I}{4 \, \pi} \, , \notag \\[1mm]
\langle j_I^\psi \rangle & = \frac{m \, a \, g^5  (c_I \, s_J \, s_K - s_I \, c_J \, c_K)}{2 \pi} \, ,
\end{align}
where the indices $I, J, K$ are never equal, while from \eqref{One_Point_Function_Def} we obtain for their scalar operator superpartners:
\begin{equation}
\langle \CO_I \rangle = \frac{m}{12 \pi} \left(-2 \, s_I^2 + s_J^2 + s_K^2 \right) \, .
\end{equation}
The holographic energy-momentum tensor and the holographic currents are conserved, 
\be
\nabla^i \langle T_{ij} \rangle =0\,, \qquad \nabla_i \langle j_I^i \rangle = 0\,,
\ee 
where $\nabla_i$ is the Levi-Civita connection of the metric on the conformal boundary, which reads
\be
\diff s^2_{\rm bdry} = - \diff t^2 + \diff s^2(S^3_{\rm bdry})\,, \quad{\rm with}\quad \diff s^2(S^3_{\rm bdry}) = \frac{1}{4g^2}\left( \sigma_1^2 + \sigma_2^2 + \sigma_3^2 \right)\,.
\ee
We can thus introduce the energy $E$ and the angular momentum $J$, defined as the conserved holographic charges associated with the Killing vectors $\frac{\partial}{\partial t}$ and $-\frac{\partial}{\partial \psi}$, respectively. These are obtained by integrating the corresponding components of the energy-momentum tensor on the boundary three-sphere $S^3_{\rm bdry}$. We find:
\begin{align}
\label{Conserved_Charges_From_Stress_Tensor}
E & = \int_{S^3_\text{bdry}} {u^i \, \langle T_{it} \rangle \, \text{vol} \left(S^3_\text{bdry} \right) } \,=\, E_0 + \frac{1}{4} \, \pi \, m \left(a^2  g^2 + 2 \, s_1^2 + 2 \, s_2^2 + 2 \, s_3^2+3 \right)   , \notag \\
J & = -\int_{S^3_\text{bdry}} {u^i \, \langle T_{i \psi} \rangle \, \text{vol} \left(S^3_\text{bdry} \right) }  \,=\,  \frac{1}{2} \pi \, a \, m \left( c_1 \, c_2 \, c_3 - s_1 \, s_2 \, s_3 \right)  ,
\end{align}
where 
\be
E_0 =  \frac{3 \, \pi }{32 \, g^2}\,,
\ee
 and we used $u = \frac{\partial}{\partial t}$ for the unit timelike vector on the conformal boundary, as well as $
\text{vol} \left(S^3_\text{bdry} \right) = \frac{1}{8g^3} \sigma_1\wedge \sigma_2\wedge \sigma_3\,.$ 
 We also obtain the conserved electric charges:
\begin{equation}
\label{Holo_Charge_From_Currents}
Q_I = \int_{S^3_{\rm bdry}}\!\!\!\! \text{vol}(S^3_{\rm bdry}) \, u_i  \langle j_I^i \rangle = - \frac{1}{16 \, \pi} \int_{S^3_{\rm bdry}}\!\!\!\! { \left( X_I^{-2} \, \star F^I + \tfrac{1}{6} C_{IJK} \, A^J \wedge F^K \right) }  =  \frac{1}{2} \, m \, \pi \, s_I  c_I \, ,
\end{equation}
where it should be noted that the Chern-Simons term evaluates to zero, implying that in this case the holographic charges are the same as the standard Maxwell charges in \eqref{QMaxwell_JKomar}.
These expressions for $E$, $J$ and $Q_I$ coincide with those obtained in \cite{Cvetic:2005zi} by other methods and reported in Section~\ref{sec:nonsusy_5d_sol}.


Next we evaluate the on-shell action. This should be computed in a regular Euclidean section of the solution. We have already described the Euclideanization and the regularity conditions to be imposed in the paragraph around eq.~\eqref{fix_gauge_5d}. Here we keep using the Lorentzian notation until the last step, taking nevertheless into account the conditions that make the Euclidean section regular.
We start from the bulk contribution, written in the form \eqref{Bulk_Action_Intermediate}. The integral over the radial coordinate is performed from the outer horizon $r_+$ up to $\rnot$.
Furthermore, applying the Stokes theorem to the second term of \eqref{Bulk_Action_Intermediate} we get:\footnote{The application of the Stokes theorem introduces a minus sign, \emph{i.e.} $\int_M{\diff (\dots)} = - \int_{\partial M}(\dots)$. This is because we chose the positive orientation to be $\diff t \wedge \diff r \wedge \sigma_1 \wedge \sigma_2 \wedge \sigma_3$. }
{\small
\begin{equation}
\label{Bulk_Action_Final}
\CS_{\text{bulk}} = - \frac{1}{12 \, \pi} \int_{M_{\rnot}}  \CV \star1 \,  + \, \frac{1}{24 \, \pi} \int_{\partial M_{\rnot}} {Q_{IJ}\, A^I \wedge \star F^J  } \, - \, \frac{1}{24 \, \pi} \int_{\partial M_{r_+}} Q_{IJ}\, A^I \wedge \star F^J  \, .
\end{equation}
}%
The first term is easily evaluated recalling the expression for $\CV$  in \eqref{Scalar_Potential_U1} and performing the bulk integral. We obtain:
\begin{equation}
\label{Bulk_Grav_Action_Computed}
- \frac{1}{12  \pi}\! \int_{M_{\rnot}}    \CV \star1  = \left[ -\frac{1}{4} \pi   g ^2  \left(\rnot^4 - r_+^4 \right) - \frac{1}{3} \, \pi   m   g ^2  \left(s_1^2+ s_2^2 + s_3^2 \right)  \left(\rnot^2 - r_+^2 \right) \right] \int{\diff t} \, ,
\end{equation}
where we displayed only the terms  that do not vanish in the limit $\rnot\to \infty$.
The terms involving $\rnot$ arise by evaluating the primitive function at the boundary, while those involving $r_+$ are the contribution of the horizon. 
In order to evaluate the second and the third terms it is convenient to fix a vielbein basis for the five-dimensional metric \eqref{Metric_CLP}. We choose:
\begin{align}
\label{Vielbein}
&e^0 = r \left(H_1  H_2 H_3 \right)^{1/6}\frac{ \sqrt Y}{\sqrt{f_1}} \, \diff t \, , \ \quad e^1 = r^2 \, \frac{\left(H_1  H_2 H_3 \right)^{1/6}}{\sqrt Y} \, \diff r \, ,\ \quad e^2 = \frac{r}{2}\left(H_1  H_2 H_3 \right)^{1/6} \sigma_1 \, ,\nn\\[1mm]
& e^3 = \frac{r}{2}\left(H_1  H_2 H_3 \right)^{1/6}  \sigma_2 \, ,\ \quad
e^4 = \frac{\sqrt{f_1}}{2 r^2 \left(H_1  H_2 H_3 \right)^{1/3}} \left(\sigma_3 - \frac{2 \, f_2}{f_1} \, \diff t \right) \, .
\end{align}
In this basis, the gauge fields read
\begin{equation}
\label{Gauge_Field_Vielbein}
A^I = D^I \, e^0 + E^I \, e^4 \, ,
\end{equation}
where:
\begin{align}
\label{Gauge_Field_Coeff}
& D^I = \left(A^I_t + 2 \, A^I_\psi \, \frac{f_2}{f_1} \right) \, \frac{\sqrt{f_1}}{r \left(H_1  H_2 H_3 \right)^{1/6} \sqrt Y} \, ,\notag \\[1mm] 
& E^I =  A^I_\psi \,\frac{2r^2\left(H_1  H_2 H_3 \right)^{1/3}}{\sqrt{f_1}} \, ,
\end{align}
while the field strengths read:
\begin{align}
F^I &= L^I \, e^{01} + M^I \, e^{14} + N^I \, e^{23} \, ,\nn\\[1mm]
\star F^I &= - L^I \, e^{234} + M^I \, e^{023} + N^I \, e^{014} \, ,
\end{align}
with
\begin{align}
& L^I = - \left[ \left(A^I_t \right)^\prime + \left(A^I_\psi \right)^\prime \frac{2 \, f_2}{f_1} \right] \, \frac{\sqrt{f_1}}{r^3 \left(H_1  H_2 H_3 \right)^{1/3}} \, , \notag \\[1mm]
& M^I = 2  \left(A^I_\psi \right)^\prime  \left(H_1  H_2 H_3 \right)^{1/6}\, \frac{ \sqrt Y}{\sqrt{f_1}} \, , \notag \\[1mm]
& N^I = - \frac{4 \, A^I_\psi}{r^2 \left(H_1  H_2 H_3 \right)^{1/3}} \, ,
\end{align}
where a prime denotes a derivative with respect to $r$.
 It follows that
\be
\left(Q_{IJ}\, A^I \wedge \star F^J \right)\big\rvert_{\partial M}   = - \frac{1}{2}\sum_{I=1}^3\, (X^I)^{-2}\left(D^I L^I + E^I  M^I \right) e^{0234} \,.
\ee
Using this formula we can evaluate the second and third term of \eqref{Bulk_Action_Final}.
It is crucial to notice that regularity of the Euclidean section requires to choose the gauge as in \eqref{fix_gauge_5d}. Doing so, we find that the horizon contribution vanishes while the boundary one gives the finite term:
\begin{equation}
\label{Bulk_Gauge_Action_Computed}
\frac{1}{24 \, \pi} \, \int_{\partial M_{\rnot}} {\left( Q_{IJ}\,{A^I} \wedge \star F^J \right) } = \frac{1}{6} \, \pi \, m \, \left[ \left(c_1 \, s_1 \, \Phi^1 + c_2 \, s_2 \, \Phi^2 + c_3 \, s_3 \, \Phi^3 \right) \right] \, \int{\diff t} \, .
\end{equation}
 The evaluation of the Gibbons-Hawking term \eqref{Gibbons-Hawking action} is straightforward and gives:
\begin{align}
\label{GH_action_computed}
\CS_{\text{GH}} & = \bigg\{ \pi \, g^2 \, \rnot^4 + 
\frac{\pi}{12} \, \left[9 + 16 \, m \, g^2 \left(s_1^2 + s_2^2 + s_3^2\right) \right] \rnot^2 \notag \\ 
& + \frac{\pi \, m}{6} \Big[- (6 + s_1^2 + s_2^2 + s_3^2)  + 
2 g^2 \left(3  a^2 + 
4   m \left(s_1^2 \, s_2^2 + s_1^2 \, s_3^2 + s_2^2 \, s_3^2 \right)\right) \Big] \bigg\} \, \int{\diff t} \, .
\end{align}
Recalling \eqref{Superpotential_Xi}, the counterterm action \eqref{Counterterm_action_simp} evaluates to:
\begin{align}
\label{Counterterm_action_computed}
\CS_{\text{ct}} & = \bigg\{ - \frac{3}{4} \, \pi \, g ^2 \, \rnot^4 + \frac{1}{4} \, \pi \, \rnot^2 \left[-4 \, m \, g ^2 \left( s_1^2 + s_2^2 + s_3^2 \right) - 3 \right] +  \, \frac{3  \pi \, m }{4}(1-a^2g^2) \notag \\
& \quad -  \pi  \,  m^2  g ^2   \left(s_1^2 \, s_2^2 + s_1^2 \, s_3^2 + s_2^2 \, s_3^2 \right)  -\frac{3 \, \pi }{32 \, g ^2} \bigg\} \, \int{\diff t} \, .
\end{align}
The regularized on-shell action $\CS_\text{reg}$ is the sum of the four terms \eqref{Bulk_Grav_Action_Computed}, \eqref{Bulk_Gauge_Action_Computed}, \eqref{GH_action_computed} and \eqref{Counterterm_action_computed}.
Adding these up, the divergences cancel out. Taking $\rnot\to \infty$ yields:
\begin{align}
\CS_\text{ren} & = \bigg\{\!-\frac{3  \pi }{32 g^2} +\frac{\pi}{12} \Big[ 2 m \left( c_1  s_1  \Phi^1 + c_2  s_2  \Phi^2 + c_3  s_3  \Phi^3 \right)  + 4 m^2  g ^2  \left(s_1^2 s_2^2 + s_1^2 s_3^2 + s_2^2 s_3^2 \right)  \notag \\[1mm]
&  \qquad   +   3 m  (g ^2  a^2 -1)  +   3 g ^2  r_+^4 + 2m\left(2g ^2  r_+^2  - 1  \right)\left(s_1^2 + s_2^2 + s_3^2 \right)  \Big] \bigg\} \int \!\diff t  \, .
\end{align}
The Euclidean action is obtained by performing the Wick rotation $t\to -i \tau$ and recalling that the Euclidean and the Lorentzian actions are related as  $\rme^{-I}=\rme^{i\CS_{\rm ren}|_{t\to -i\tau}}$ in the gravitational path integral. Effectively this means that we just have to replace $\int\diff t \to -\int \diff \tau$ in the expression above. As usual, regularity of the Euclidean solution as $r\to r_+$ fixes the circumference of the Euclidean time circle to be $\int \diff \tau = \beta$, where $\beta$ is the inverse Hawking temperature given in~\eqref{Entropy_Temperature_Potentials}.
 In this way we reach the result reported in \eqref{On_Shell_Action}.

\section{The on-shell action of AdS$_4$ black holes}\label{app:OnShActionsAdS4}

In this appendix, we evaluate the on-shell action and the holographic charges of  the four-dimensional solution in Section~\ref{4d_section}. As in the five-dimensional case, we use the method of holographic renormalization.

The four-dimensional metric has the same form as \eqref{Induced_Metric}, where now $i,j = 0, \dots, 2$.
The renormalized action is again $\CS_{\text{ren}} = \lim_{\rnot \to \infty}  \, \CS_{\text{reg}}$ 
with
$\CS_{\text{reg}} = \CS_{\text{bulk}} + \CS_{\text{GH}} + \CS_{\text{ct}} \, .$
Using the Einstein equation, the bulk supergravity action \eqref{Lagrangian_Four_Dim} can be recast into
\begin{align}
\label{Bulk_Action_Recasted}
\CS_\text{bulk}  = - \frac{1}{16\pi} & \int_{M_{\rnot}} { \Big(   - 2 \CV\star 1 - \frac{1}{2} \rme^{-\scal} F_{3}\wedge \star F_3    - \frac{1}{2 \left(1 + \chi^2  \,\rme^{2 \scal} \right) } \rme^{\scal} F_{1}\wedge \star F_1 } \notag \\[1mm]
& \ \qquad - \frac{1}{2} \, \chi \, F_3 \wedge F_3 + \frac{\chi \, \rme^{2 \scal}}{2 \left(1 + \chi^2 \, \rme^{2 \scal} \right)  } F_1 \wedge F_1    \Big) \,.  
\end{align}
The Gibbons-Hawking boundary integral is defined as in five dimensions,
\begin{equation}
\label{Gibbons-Hawking action_Four_Dim}
\CS_{\text{GH}} = \frac{1}{8 \, \pi} \, \int_{\partial M_{\rnot}} {\diff^3 x \, \sqrt{h} \, K} \, .
\end{equation}
The counterterm action reads
\begin{align}
\label{Counterterm_action_Four_Dim}
\CS_\text{ct} = - \frac{1}{8 \, \pi} \int_{\partial M_{\rnot}} \diff^3 x \, \sqrt{h} \, \CW \left( 1 + \frac{1}{4 \, g^2} \, R  \right) \, ,
\end{align}
where the real superpotential reads
\begin{equation}
\label{W_function_four_dim}
\CW = g \, \rme^{\scal/2} \sqrt{\chi^2+\left(\rme^{-\scal}+1\right)^2} \, .
\end{equation}
We have obtained this superpotential by specializing the results of \cite{Cabo-Bizet:2017xdr} to our case. This reference derived the holographic counterterms for Fayet-Iliopoulos U(1)$^4$ supergravity, that is four-dimensional $\mathcal{N}=2$ supergravity coupled to three vector multiplets and with an Abelian gauging of the R-symmetry. This is related to the theory considered in the present paper by setting the four gauge fields pairwise equal, and the same for the scalar symplectic sections, 
 $X^0=X^1= (\rme^{-\scal} + i \chi)^{-1/2}$, $X^2=X^3=(\rme^{-\scal} + i \chi)^{1/2}$. It should be noted that the counterterm \eqref{Counterterm_action_Four_Dim} is compatible with supersymmetry provided a combination of the scalar fields is given Neumann boundary conditions~\cite{Freedman:2013ryh}; this means that our renormalized action is a function of vevs for the operators dual to these scalars, and of sources for the other operators~\cite{Cabo-Bizet:2017xdr}.

We now evaluate the terms above on the solution. Displaying only the contributions that do not vanish in the limit $\rnot\to \infty$, the bulk action \eqref{Bulk_Action_Recasted} yields
\begin{align}
	\label{Bulk_Action_Evaluated_Four_Dim}
\CS_\text{bulk}   = \frac{\int \diff t }{2(a^2 g^2-1)}\bigg\{&g^2 (\rnot^3- r_+^3) \, + \, 3 g^2  m (\rnot^2-r_+^2) (s_1^2+ s_2^2)  \nn\\[1mm]
&+   (\rnot- r_+) \left[a^2g^2+2\, m^2g^2 (s_1^4 + 4 \, s_1^2 s_2^2+ s_2^4 )\right] \notag \\[1mm]
& -\frac{2\,m^2 \left[ c_1^2 \, s_1^2 \left(2 \, m \, s_2^2 + r_+ \right) + c_2^2 \, s_2^2 \left(2 \, m \, s_1^2 + r_+ \right)\right]}{ a^2+\left(2 \, m \, s_1^2 + r_+ \right) \left(2 \, m \, s_2^2 + r_+ \right)} \bigg\}  \, ,
\end{align}
the Gibbons-Hawking term gives:
\begin{align}
\label{Gibbons-Hawking_Evaluated_Four_Dim}
\CS_\text{GH}  = \frac{\int\diff t}{2 \left(1 - a^2 g^2\right)}\bigg\{ & 3  g^2 \rnot^3 + 9 m g^2   \rnot^2 (s_1^2 + s_2^2 ) + \left[ \tfrac{5}{3}  a^2 g^2+6  m^2g^2 (s_1^4 + 4  s_1^2  s_2^2 + s_2^4 ) + 2\right]  \rnot  \notag \\[1mm]
&  +  m \left(\tfrac{5}{3} \, a^2 g^2 +12 \,   m^2g^2  s_1^2  s_2^2  - 1 \right) \left(s_1^2+ s_2^2 \right)-3m \bigg\} \, ,
\end{align}
while the counterterm action evaluates to:
\begin{align}
\label{Counterterm_Evaluated_Four_Dim}
\CS_\text{ct}  = \frac{\int{\diff t} }{1 - a^2 g^2}\bigg\{ & -g^2  \rnot^3 -3  g^2  m \, \rnot^2 \left(s_1^2 + s_2^2 \right) -\left[\tfrac{1}{3} a^2 g^2 + 2   m^2g^2 \left(s_1^4 + 4  s_1^2  s_2^2 + s_2^4 \right) + 1 \right]  \rnot  \notag \\[1mm]
&  -  m  \left(\tfrac{1}{3} a^2 g^2 + 4 \,   m^2g^2  s_1^2  s_2^2  \right) \left(s_1^2 + s_2^2 \right)+m  \bigg\}  \, .
\end{align}
Adding up these three expressions and sending $\rnot \to \infty$, we obtain our result for the renormalized action:
\begin{align}
\CS_\text{ren} & = \frac{\int{\diff t}}{2(1-a^2g^2)}\, \bigg\{  g^2r_+^3 + 3 \, m \,g^2 r_+^2 \left(s_1^2 + s_2^2 \right) + r_+ \left[a^2g^2 + 2 \, m^2g^2 \left(s_1^4 + 4  s_1^2  s_2^2 + s_2^4 \right) \right]  \nn\\[1mm]
&\quad\qquad\qquad\qquad  + m  \left(a^2g^2 + 4 \, m^2g^2 s_1^2 s_2^2 -1 \right) (s_1^2+s_2^2)  - m   \notag \\[1mm]
&\quad\qquad\qquad\qquad + \frac{2\,m^2 \left[ c_1^2  s_1^2 \left(2 \, m \, s_2^2 + r_+ \right) + c_2^2 s_2^2 \left(2 \, m \, s_1^2 + r_+ \right)\right]}{ a^2+\left(2 \, m \, s_1^2 + r_+ \right) \left(2 \, m \, s_2^2 + r_+ \right)} \,  \bigg\}  \, .
\end{align}
The Euclidean on-shell action $I$ is obtained by Wick-rotating $t=-i\tau$ and identifying $\tau\sim\tau+\beta$, where $\beta$ was given in \eqref{Properties_Solution_Four_Dim}. Differently from the five-dimensional case, there is no subtlety related to the choice of a regular gauge, because the four-dimensional action is gauge-invariant. Therefore one simply has $I = -i \CS_\text{ren}|_{\int\diff t \to -i\beta}$
 Our final result is displayed in \eqref{On_Shell_Action_Euclidean_Four_Dim}.

The holographic energy-momentum tensor is given by:
\begin{align}
\langle T_{ij} \rangle &= - \lim_{\rnot \to \infty} \frac{2 \, \rnot \, g}{\sqrt{h}} \, \frac{\delta \CS_\text{reg}} {\delta h^{ij}}   \nn\\[1mm]
 &= - \frac{1}{8 \, \pi} \, \lim_{\rnot \to \infty} \rnot \, g \left[ K_{ij} - \left(K - \CW \right) h_{ij} - \frac{1}{2 \, g^2} \, \CW  \left(R_{ij} - \frac{1}{2} \, R \, h_{ij} \right)  \right] \, .\label{Stress_Energy_Tensor_Formula_Four_Dim}
\end{align}
The charges appearing in~\eqref{Energy, Angular Momentum, Charges_Four_Dim} are evaluated in a frame which is non-rotating at infinity, so in order to compare with those expressions it is convenient to use the time and angular coordinates $t',\phi'$ defined in \eqref{Change_To_Non_Rotating}. Here we report only the components $\langle T_{t't'} \rangle$ and $\langle T_{t' \phi'} \rangle$, since these are the only ones needed to compute the energy and the angular momentum: 
\begin{align}
\label{Stress_Energy_Tensor_Components_Four_Dim}
\langle T_{t't'} \rangle & = \frac{g^2  m \left(s_1^2 + s_2^2 + 1\right) \left(1-a^2g^2\cos^2\theta\right)(2-2a^2g^2\cos^2\theta+a^2g^2\sin^2\theta)}{8 \, \pi  \left(a^2  g^2 - 1 \right)^2} \, , \notag \\[1mm]
\langle T_{t' \phi'} \rangle & = \frac{3 \, a \, g^2  m  \left(s_1^2 + s_2^2 + 1\right) \,\sin^2{\theta }\left(a^2 g^2 \cos^2 \theta - 1 \right)}{8 \, \pi  \left(a^2 g^2-1\right)^2}  \, .
\end{align}
 The asymptotic metric at $r \to \infty$ is
\begin{equation}
\diff s^2 = \frac{\diff r^2}{g^2  r^2} + g^2  r^2 \, \diff s_{\text{bdry}}^2  \, ,
\end{equation}
where the metric on the conformal boundary reads in the non-rotating frame\begin{equation}
\label{Boundary_Metric_Four_Dimensions}
\diff s_{\text{bdry}}^2 = -\frac{\Delta_\theta}{\Xi} \, \diff t'^2 + \frac{\diff \theta^2}{g^2 \Delta_\theta }  + \frac{\sin ^2 \theta \, \diff \phi'^2}{g^2 \Xi }  \, ,
\end{equation}
and $\Delta_\theta,\Xi$ were given in \eqref{Quantites_Four_Dimensional}.\footnote{The metric \eqref{Boundary_Metric_Four_Dimensions} is related by a Weyl transformation and a change of coordinate to the canonical metric on $\mathbb{R}\times S^2$:
$
\frac{\Xi}{\Delta_\theta}\diff s^2_{\rm bdry} = -\diff t^2 + \frac{1}{g^2}\left( \diff \theta'{}^{\,2} + \sin^2\theta'\,\diff \phi'^{\,2}\right)$,
with $\tan\theta = \sqrt{1-a^2g^2}\,\tan\theta'$. We will not need to implement this transformation here.
}
Using these expressions we can evaluate the conserved charges $E$ and $J$, associated with the symmetries generated by $\frac{\partial}{\partial t'}$ and $-\frac{\partial}{\partial \phi'}$, respectively.
We obtain:
\begin{align}
E &= \int_{\Sigma_\text{bdry}} {u^i \, \langle T_{it'} \rangle \, \text{vol} \left(\Sigma_\text{bdry} \right) } = \frac{m}{\Xi^2} \left(1 + s_1^2 + s_2^2 \right)   \, ,\nn\\[1mm]
J &= -\int_{\Sigma_\text{bdry}} {u^i \, \langle T_{i \phi'} \rangle \, \text{vol}\left(\Sigma_\text{bdry} \right) }  =  \frac{a \, m}{\Xi^2} \left(1 + s_1^2 + s_2^2 \right)\, .
\end{align}
where $u =   \sqrt{\frac{\Xi}{\Delta_\theta} }\, \frac{\partial}{\partial t'}$ is the unit, outward-pointing timelike vector and $\Sigma_\text{bdry}$ is the two-dimensional Cauchy surface at the boundary, with metric induced from \eqref{Boundary_Metric_Four_Dimensions}.
These expressions coincide with the ones computed in \cite{Chong:2004na} and reported in~\eqref{Energy, Angular Momentum, Charges_Four_Dim}. The electric charges obtained from the holographic currents $\langle j^i\rangle$ also agree with those in~\eqref{Energy, Angular Momentum, Charges_Four_Dim}.

\section{Legendre transform of general AdS$_5$ black hole entropy function}\label{app:Leg_transf}

In this appendix, we prove that the Legendre transform of the entropy function \eqref{entropyfctSE} leads to the entropy of the asymptotically AdS$_5$ BPS black holes of \cite{Gutowski:2004yv,Kunduri:2006ek}. These are solutions to five-dimensional Fayet-Iliopoulos gauged supergravity coupled to $n_V$ vector multiplets, which carry angular momenta $J_1^\star,J^\star_2$ and $n_V+1$ electric charges $Q_I^\star$. 
 The proof is a generalization to Fayet-Iliopoulos gauged supergravity of the procedure presented in \cite[Appendix B]{Cabo-Bizet:2018ehj} for the $\U(1)^3$ theory. Although straightforward, this requires extensive use of the properties of the tensor $C_{IJK}$ and for this reason we discuss it in some detail.

We start from the function of the rotational and electric chemical potentials $\omega_i$, $i=1,2$, and $\varphi^K$, $K=1,\ldots,n_V+1$, proposed in \cite{Hosseini:2018dob}:
 \be
 I = \frac{\pi}{24}\,\frac{C_{IJK}\varphi^I \varphi^J \varphi^K}{\omega_1\,\omega_2}\,. 
 \ee
 We want to compute the Legendre transform, subject to the constraint
 \be\label{gen_constr_app}
\omega_1 + \omega_2 - 3\bar{X}_K \varphi^{K} =  2\pi i n\,,
\ee
where  $n$ is a real number, and we recall from Appendix~\ref{app:OnShActionsAdS5} that $\bar{X}_K$ are the values taken by the scalar fields in the supersymmetric AdS$_5$ solution, which can be traded for the Fayet-Iliopoulos gauging parameters.
Following \cite{Cabo-Bizet:2018ehj}, we set up the extremization problem
 \be\label{def_Legendre_transf}
 {S}(Q_K,J_i) =  {\rm ext}_{\{\varphi^K,\,\omega^i,\,\Lambda\}}\big[- {I} - \varphi^K Q_K - \omega^i J_i - \Lambda(  \omega_1 + \omega_2 - 3\bar{X}_K \varphi^K - 2\pi i n )\big]\, ,
\ee
where $\Lambda$ is a Lagrange multiplier implementing the constraint, and $\omega^i=\omega_i$.  The extremization equations obtained from \eqref{def_Legendre_transf} are
\begin{align}\label{extrem_eqs}
-\frac{\partial I}{\partial \varphi^K} = Q_K - 3 \bar{X}_K \,\Lambda\ ,\qquad\qquad -\frac{\partial I}{\partial \omega^i} = J_i + \Lambda\ ,
\end{align}
together with the constraint \eqref{gen_constr_app} that follows from varying with respect to $\Lambda$. Using the identity \eqref{3Cinto1C}, it is not hard to see that the equations above imply
\be
0 = \frac{1}{6}\,C^{IJK}(-Q_I + 3 \bar{X}_I \,\Lambda)(-Q_J + 3 \bar{X}_J \,\Lambda)(-Q_K + 3 \bar{X}_K \,\Lambda) - \frac{\pi}{4} (J_1+\Lambda) (J_2+\Lambda) \,.
\ee
 This can be written as
\be\label{cubiceq_Lambda}
0=\, p_0 + p_1 \Lambda + p_2 \Lambda^2 + \Lambda^3\ ,
\ee
with
\begin{align}\label{p0p1p2_FIsugra}
p_0 &= -\frac{1}{6}\,C^{IJK}Q_IQ_JQ_K - \frac{\pi}{4} J_1J_2\ ,\nn\\[1mm]
p_1 &=  \frac{3}{2}\, C^{IJK} \bar{X}_I Q_JQ_K -\frac{\pi}{4}\,( J_1 + J_2) \ ,\nn\\
p_2 &= -\frac{9}{2}\, C^{IJK} \bar{X}_I \bar{X}_JQ_K -\frac{\pi}{4}\ .
\end{align}
The cubic equation \eqref{cubiceq_Lambda} is straightforwardly solved for $\Lambda$, and we denote by ``Roots''  the set of three solutions.
We can also solve the rest of the equations \eqref{extrem_eqs}, together with the constraint \eqref{gen_constr_app}, and in this way determine the saddle point values of the chemical potentials $\omega^i,\varphi^I$. 
These read:
\begin{align}\label{saddles_chemical}
& \omega^i = \frac{1}{6}\,\Xi\,C^{IJK}\tilde Q_I\tilde Q_J \tilde Q_K  |\epsilon^{ij}|\tilde J_j   \ , \qquad \varphi^I  =  - \frac{1}{2}\,\Xi\,C^{IJK}\tilde Q_J \tilde Q_K \tilde J_1 \tilde J_2 \ ,
\end{align}
where we introduced
$\tilde Q_I = Q_I -3\bar{X}_I\Lambda$, $\tilde J_i = J_i +\Lambda$, along with
\be\label{def_Xi}
\Xi = \frac{4\pi i n}{3\tilde J_1 \tilde J_2 \, C^{IJK}\bar{X}_I \tilde Q_J \tilde Q_K+ \frac{1}{3}(\tilde J_1 + \tilde J_2) C^{IJK}\tilde Q_I \tilde Q_J \tilde Q_K}\ ,
\ee
and it is understood that $\Lambda \in {\rm Roots}$.

Now the same argument used in \cite{Cabo-Bizet:2018ehj} shows that the Legendre transform  reads
\be\label{S_supLambda}
 {S} = {\rm ext}_{\Lambda\,\in\,{\rm Roots}} \,(2\pi in\,\Lambda)\ ,
\ee
and that $S$ is real and positive if and only if one imposes
\be\label{p0isp1p2}
p_0 = p_1p_2\,,
\ee
and picks the purely imaginary root $\Lambda = i\sqrt{p_1}$ if $n<0$, or $\Lambda =-i\sqrt{p_1}$ if $n>0$. Recalling \eqref{p0p1p2_FIsugra}, we see that \eqref{p0isp1p2} is a constraint on the charges.
Assuming this condition, the Legendre transform \eqref{S_supLambda} reads 
\begin{align}\label{entropy_p1}
 {S} &= 2\pi| n |\,\sqrt{p_1} \nn\\[1mm]
 &= \pi |n|\sqrt{ 6\, C^{IJK} \bar{X}_I Q_JQ_K -\pi\,( J_1 + J_2) } \ .
\end{align}
For $n= \mp 1$, this is the Bekenstein-Hawking entropy of the black holes of \cite{Gutowski:2004yv,Kunduri:2006ek}, in the form first given in \cite{Kim:2006he}. This concludes our proof.

Specializing to the $\U(1)^3$ model, which as we reviewed in  Appendix \ref{app:OnShActionsAdS5} implies taking  $C_{123}=1$ and $\bar{X}_I = \frac{1}{3}$,
 all the expressions above reduce to those given in \cite[Appendix B]{Cabo-Bizet:2018ehj}, upon identifying $Q_I^{\rm here}=-Q_I^{\rm there}$, $\varphi^I_{\rm here}=-\Delta_{I\,\rm there}$ and $\mu_{\rm there}=-\frac{\pi}{4}$. Moreover, for $n=\mp 1$ the constraint \eqref{gen_constr_app} is perfectly consistent with the one we obtained in Section~\ref{5d_section} analyzing the black hole solution to the $\U(1)^3$ model.

For the multi-charge solution of \cite{Gutowski:2004yv}, where the two angular momenta are set equal, 
we can show that the explicit expressions for the chemical potentials we gave in \eqref{conjecture} satisfy the saddle point expressions \eqref{saddles_chemical}.
The solution of  \cite{Gutowski:2004yv} is controlled by real parameters
\be
q_I\,, \quad\alpha_1=\frac{27}{2}C^{IJK}\bar X_I \bar X_J q_K\,,\quad \alpha_2= \frac{27}{2}C^{IJK}\bar X_I q_J q_K\,,\quad \alpha_3=\frac{9}{2}C^{IJK}q_Iq_Jq_K\,.
\ee
The explicit formulae for the BPS black hole entropy and charges in terms of these parameters are \cite{Gutowski:2004yv}:
 \begin{align}\label{BPS_charges_GR2}
 S^\star &= \frac{1}{4}\pi^2\sqrt{4 \left(1+\alpha_1 \right)\alpha_3-\alpha_2^2}\,,\qquad   J_1^\star=J_2^\star \equiv J^\star = \frac{\pi}{8}\left(\alpha_2+2\alpha_3 \right) \,,\nn\\[1mm]
 Q^\star_I &= \frac{3}{4}\,\pi \left( q_I - \tfrac{1}{2}\alpha_2 \bar{X}_I + \tfrac{3}{2} C_{IJK} \bar{X}^J C^{KLM} q_Lq_M\right)\,.
\end{align}
 The charges satisfy the relation \eqref{p0isp1p2};
correspondingly, the BPS solution has just $n_V+1$ independent parameters, though there are $n_V+2$ charges $J^\star, Q_I^\star$.
In order to compare \eqref{saddles_chemical} with \eqref{conjecture}, we need to evaluate $C^{IJK} \tilde Q_J \tilde Q_K$, $C^{IJK}\tilde Q_I \tilde Q_J \tilde Q_K$, $C^{IJK}\bar{X}_I \tilde Q_J \tilde Q_K$ in terms of the parameters. Straightforward though tedious computations making repeated use of the identity \eqref{3Cinto1C} lead us to:\footnote{These expressions also hold for the solution of \cite{Kunduri:2006ek}, where the two angular momenta are different.}
\begin{align}\label{contractions_tQ}
C^{IJK} \tilde Q_J^\star \tilde Q_K^\star &= \frac{1}{8}\left[ \pi^2\left((1+\alpha_1)\alpha_3 - \tfrac{1}{4}\alpha_2^2\right) +16 \,\Lambda^2 \right]\bar{X}^I \nn\\[1mm]
&\quad+ \frac{9}{16} \left[  \pi^2\left(1+\alpha_1+\tfrac{1}{2}\alpha_2\right) -4\pi\Lambda \right] C^{IJK}q_Jq_K \nn\\[1mm]
&\quad -\frac{9}{16}\left[ \pi^2 \left(\alpha_2+2\alpha_3\right) +8\pi\Lambda \right] C^{IJK}\bar{X}_Jq_K \,, \nn\\[1mm]
C^{IJK}\tilde Q_I^\star \tilde Q_J^\star \tilde Q_K^\star &= \frac{3\pi^3}{32}\!\left( \left(1+2\alpha_1+\alpha_1^2-\alpha_3\right)\alpha_3 +\tfrac{1}{2}(\alpha_1-1)\alpha_2\alpha_3 - \tfrac{1}{4}(2+\alpha_1) \alpha_2^2 - \tfrac{1}{8}\alpha_2^3\right)
\nn\\[1mm]
& \quad - \frac{3\pi^2}{8} \left(  \alpha_1\alpha_3 + \alpha_2+3 \alpha_3 -  \tfrac{1}{4}\alpha_2^2\right)\Lambda  +\frac{3\pi}{4} (2\alpha_1+\alpha_2)\Lambda^2 -6\Lambda^3\,,    \nn\\[1mm]
C^{IJK}\bar{X}_I \tilde Q_J^\star \tilde Q_K^\star &= \frac{\pi^2}{24}\left( \alpha_1\alpha_3+\alpha_2+3\alpha_3 - \tfrac{1}{4}\alpha_2^2 \right) -\frac{\pi}{6}\left( 2\alpha_1+\alpha_2 \right) \Lambda + 2\Lambda^2 \,.
\end{align}
We are interested in the case where the Legendre transform is real and positive, and $n=\mp1$. We thus substitute $\Lambda = \pm i\sqrt{p_1} = \pm  \frac{i}{2\pi}S^\star$ in the formulae above. Note that then the term proportional to $\bar{X}^I$ in the first line of \eqref{contractions_tQ} vanishes.
Plugging \eqref{BPS_charges_GR2}, \eqref{contractions_tQ} into \eqref{saddles_chemical}, \eqref{def_Xi}, we obtain precisely the expressions for the chemical potentials given in \eqref{conjecture}.
As a consistency check, we also verified that \eqref{entropyfctSE}, \eqref{conjecture}, \eqref{BPS_charges_GR2} satisfy the supersymmetric quantum statistical relation $I^\star = - S^\star -  \omega^\star J^\star -  \varphi^{I\star}  Q^\star_I$.


\bibliography{BPSlimitBHthermo}
\bibliographystyle{JHEP}

\end{document}